\documentstyle[a4,amsmath,amssymb,latexsym,epsfig,citesort]{article}
\def \d{{\mathrm{d}}}

\def \pd{\partial}

\def \stimes{\times\!\!\!\!\!\!\supset}

\def \tl#1{\overset{\kern 1pt\circ}{#1}}
\def \TL#1{\overset{\kern -3pt \circ}{#1}}
\def \TLL#1{\overset{\kern -7pt \circ}{#1}}

\def \burger{{\boldsymbol{b}}}
\def \frank{\boldsymbol{\Omega}}

\begin{document}
\title{{\bf Twist disclination in the field theory of elastoplasticity}}
\author{Markus Lazar\thanks{Present address: 
	Laboratoire de Mod{\'e}lisation en M{\'e}ca\-nique,
        Universit{\'e} Pierre et Marie Curie, Tour 66,
	4~Place Jussieu, Case 162, F-75252 Paris C{\'e}dex 05, France.}\\
        Max-Planck-Institute for Mathematics in the Sciences,\\
        Inselstr. 22-26, D-04103 Leipzig, Germany\\
        E-mail: lazar@mis.mpg.de}

\date{\today}    
\maketitle
\begin{abstract}
In this paper we study the twist disclination within the elastoplastic defect theory. 
Using the stress function method, we found 
exact analytical solutions for all characteristic fields
of a straight twist disclination in an infinitely extended linear isotropic medium. 
The elastic stress, elastic strain and displacement  
have no singularities at the disclination line.  
We found modified stress functions for the twist disclination.
In addition, we calculate the disclination density, 
effective Frank vector, disclination torsion and effective Burgers vector
of a straight twist disclination. By means of gauge theory of defects
we decompose the elastic distortion into the translational and rotational
gauge fields of the straight twist disclination.
\\

\noindent
{\bf Keywords:} disclinations; dislocations; gauge theory of defects; stress functions\\
{\bf PACS:} 61.72.Lk, 62.20.-x, 81.40.Jj
\end{abstract}
\vspace*{5mm}

\section{Introduction}
\setcounter{equation}{0}
Disclinations are very important and interesting lattice defects. 
They may be defects in warped and twisted materials.
Disclinations have been investigated in the context of applications
to liquid crystals as twisting discontinuities~\cite{Frank,Kleman},
Abrikosov lattices formed by magnetic flux lines in the mixed state of type-II
superconductors~\cite{TE68}, polymers by chain kinking and twisting of molecules~\cite{LG70}, 
Bloch wall lattices~\cite{HW70,Klemanb}, 
biological structures~\cite{Harris}, amorphous bodies~\cite{Richter}
and rotation plastic deformations~\cite{RV83,RV}.
Because disclinations cause strong elastic distortions and lattice bending 
it seems that very strong distortions are necessary in order to realize disclinations 
in crystals.

A disclination is characterized by a closure failure of the rotation
for a closed circuit round the disclination line.
There are wedge and twist disclinations.
If the Frank angle (rotation failure) of the disclination is a symmetry angle of lattice, 
then the disclination is called a perfect disclination. 
Such disclinations have been introduced by Anthony~\cite{Anthony70} and deWit~\cite{deWit73a,deWit73b}.
In the case of a twist disclination the rotation axis is perpendicular to the disclination line.
The smallest value of the Frank vector is 
$\pi/2$ in a cubic lattice and $\pi/3$ in a hexagonal lattice. 
If the Frank angle is not a symmetry angle of lattice, the disclination is 
called partial disclination. 
They play an important role, e.g., in building 
of twin boundaries (see, e.g.,~\cite{deWit72}).
Disclinations correspond, in general, to Volterra's distortions of the second kind 
(see also~\cite{KA75}). 
Thus, these defects are of rotational type. 
They are different from the so-called Frank's (spin) disclinations which are 
elementary defects in liquid crystals (see~\cite{Frank}).

The traditional description of elastic fields produced by defects
(e.g. dislocations, disclinations and cracks)
is based on the classical theory of linear elasticity.
However, classical elasticity breaks down near the defect line and leads 
to singularities. This is unfortunate since the defect core
is a very important region in the theory of defects.
Of course, such singularities
are unphysical and an improved model of defects should eliminate them.

On the other hand, there are other non-standard continuum models of 
defects, e.g., the nonlocal continuum model~\cite{Eringen83,Eringen85,Eringen87,Eringen02,Pov,PM00}, 
the strain gradient elasticity~\cite{GA96,GA97,GA99,Aifan99,GA99b,GA00,Gutkin00} 
and the field theory of elastoplasticity which has been developed from the
gauge theory of defects~\cite{Lazar00,Lazar02a,Lazar02b,Lazar02c,Lazar02d,Lazar03}. 
All these theories are successfully applied to the description of screw and edge
dislocations. In this context the stresses have no singularities at the dislocation line.
In addition, the dislocation core arises naturally.
In particular, the field theory of elastoplasticity is a gauge theory of defects in which
the defects cause plasticity. The corresponding gauge fields may 
be identified with the plastic distortion.
By the help of this theory the elastic and plastic part
of the total distortion can be calculated. 
The total distortion is defined in terms of a displacement and consists of the
elastic and plastic part.
In the case of dislocations~(see, e.g.,~ \cite{Lazar02d})
the elastic distortion is continuous even in the dislocation core
and the plastic part becomes discontinuous.
But in the case of disclinations the situation is less complete worked out.
The stresses of straight wedge and twist disclinations have been calculated 
by Povstenko~\cite{Pov} in the framework of
Eringen's nonlocal elasticity and by Gutkin and Aifantis~\cite{GA99b,GA00,Gutkin00} 
by the help of strain gradient elasticity. 
However, no rotation and displacement vectors, no bend-twist and 
no disclination and dislocation density tensors were obtained in their works. 
In a recent paper~\cite{Lazar03} the wedge disclination has been 
investigated in the field theory of elastoplasticity. It was possible to 
calculate all characteristic field quantities. It has been seen that the disclination core may be defined
quite natural in this framework.

In this paper we want to extend our study for a 
straight twist disclination.
We use the field theory of elastoplasticity 
to find nonsingular solutions for the stress and strain fields
and the rotation 
and displacement fields. In addition, we investigate the relation to gauge theory of defects.
We use the stress function method and hope to close
the gap between the non-local and strain gradient results for the 
case of a straight twist disclination.
In this framework we want to work out all geometric quantities of a twist disclination.

\section{Basic equations}
\setcounter{equation}{0}
In this section we apply the field theory of elastoplasticity to
the case of a straight twist disclination.
In elastoplasticity the elastic distortion is given by~\cite{Lazar00,Lazar02a,Lazar02b,Lazar02c}
\begin{align}
\label{dist2}
\beta_{ij}=\pd_j u_i+\tilde\beta_{ij}.
\end{align}
It is an additive decomposition of the elastic distortion into compatible and 
purely incompatible distortion. 
This decomposition can be justified by the help of the gauge 
theory of defects~\cite{Lazar00}.
The displacement field $u_i$ gives rise to a compatible distortion 
and the tensor $\tilde\beta_{ij}$ is the proper incompatible part of 
the elastic distortion.

The Burgers vector $b_i$ is defined by the help of the distortion tensor
\begin{align}
\label{Burgers}
b_i=\oint_\gamma\beta_{ij}\d x_j,
\end{align}
where $\gamma$ denotes the Burgers circuit.
In elastoplasticity the linear elastic strain tensor 
is given by means of the incompatible distortion tensor~(\ref{dist2})
according to
\begin{align}
E_{ij}\equiv\beta_{(ij)}=\frac{1}{2}\big(\pd_i u_j+\pd_ju_i+\tilde\beta_{ij}+\tilde\beta_{ji}\big),
\qquad E_{ij}=E_{ji}.
\end{align}

The force stress is the response quantity to elastic strain and 
is given by the (generalized) Hooke's law
for an isotropic medium
\begin{align}
\sigma_{ij}=
2\mu\left( E_{ij}+\frac{\nu}{1-2\nu}\,\delta_{ij} E_{kk}\right),
\qquad\sigma_{ij}=\sigma_{ji},
\end{align}
where $\mu$, $\nu$ are shear modulus and Poisson's ration, respectively.
The force stress satisfies the force equilibrium condition
\begin{align}
\label{FEq}
\pd_j\sigma_{ij}=0.
\end{align}
The inverse of Hooke's law reads
\begin{align}
\label{Hooke-inv}
E_{ij}=
\frac{1}{2\mu}\left( \sigma_{ij}-\frac{\nu}{1+\nu}\,\delta_{ij} \sigma_{kk}\right).
\end{align}

In the conventional disclination theory~\cite{Anthony70,Mura72,deWit73a,deWit73b}
 the torsion tensor (linear version of Cartan's torsion)
is defined by
\begin{align}
\label{tors1}
\alpha_{ij}:=
\epsilon_{jkl}\big(\pd_k\beta_{il}+\epsilon_{ilm}\varphi^*_{mk}\big)
=\epsilon_{jkl}\big(\pd_k\tilde\beta_{il}+\epsilon_{ilm}\varphi^*_{mk}\big).
\end{align}
Anthony called it the disclination torsion (see~\cite{Anthony70}). 
On the other hand, 
it is sometimes called dislocation density in the theory of disclinations 
(see, e.g., \cite{Mura72,deWit73a,deWit73b}). 
The $\varphi^*_{ij}$ was introduced by Mura~\cite{Mura72}
as ``plastic rotation'' and deWit~\cite{deWit73a,deWit73b,deWit72b} called
this quantity ``disclination loop density''. The field $\tilde\beta_{ij}$ 
may be identified with deWit's ``dislocation loop density''.
For a dislocation it yields $\varphi^*_{ij}=0$ and then (\ref{tors1}) 
has the shape of a proper dislocation density.
Using the elastic  bend-twist tensor (see, e.g.,~\cite{Kosseka})
\begin{align}
\label{bt}
k_{ij}=\pd_j\omega_i-\varphi^*_{ij},
\end{align}
with the rotation vector 
\begin{align}
\omega_i=-\frac{1}{2}\,\epsilon_{ijk}\beta_{jk},
\end{align}
Eq.~(\ref{tors1}) can be rewritten according to (see also~\cite{Anthony70,deWit72b,Kosseka})  
\begin{align}
\label{DislD1}
\alpha_{ij}=
\epsilon_{jkl}\big(\pd_k E_{il}+\epsilon_{iml} k_{mk}\big)
=\epsilon_{jkl}\pd_k E_{il}+\delta_{ij} k_{ll}-k_{ji}.
\end{align}
The index $i$ indicates the direction of the Burgers vector, 
$j$ the dislocation line direction. 
Thus, the diagonal components of $\alpha_{ij}$ represent screw dislocations,
the off-diagonal components edge dislocations.

The so-called disclination density tensor of a discrete disclination
is defined by~\cite{Anthony70,deWit73a,deWit73b,Mura72,deWit72b,Kosseka}
\begin{align}
\label{DisclD1}
\Theta_{ij}:=\epsilon_{jmn}\pd_m k_{in}=-\epsilon_{jmn}\pd_m \varphi^*_{in}.
\end{align}
The index $i$ indicates the direction of the Frank vector, 
$j$ the disclination line direction. 
Thus, the diagonal components of $\Theta_{ij}$ represent wedge disclinations,
the off-diagonal components twist disclinations.
The Frank vector $\Omega_i$ is defined by the help of the elastic 
bend-twist tensor
\begin{align}
\label{Frank}
\Omega_i=\oint_\gamma k_{ij}\d x_j.
\end{align}

Consequently, the dislocation density and the disclination density satisfy the 
following compatibility equations (1st and 2nd Bianchi identities)
\begin{align}
\label{1Bianchi}
\pd_j\alpha_{ij}&-\epsilon_{ikl}\Theta_{kl}=0,\\
\label{2Bianchi}
&\pd_j\Theta_{ij}=0.
\end{align}

On the other hand, 
the theory of defects (dislocations and disclinations) 
can be considered as a gauge model of defects in solids~\cite{KE83,EL88}.
The gauge group is the group $ISO(3)=T(3)\stimes SO(3)$ ($T(3)$ -- 
three-dimensional translational group, $SO(3)$ -- three-dimensional 
rotational group and $\stimes$ denotes the semi-direct product).
In this framework, we are able to decompose the incompatible distortion~(\ref{dist2}).
Namely, the incompatible distortion takes the (linearized) form~\cite{KE83,EL88,Maly96,Maly00}
\begin{align}
\label{dist-GT}
\tilde\beta_{ij}=\phi_{ij}+\epsilon_{ikl} W_{kj} x_l,
\end{align}
where $\phi_{ij}$ and $W_{ij}$ are the translational  and rotational gauge 
fields, respectively. 
More precisely, $\phi_{ij}$ is the translational part of the 
generalized affine connection~\cite{MAG,Mielke} 
and $W_{ij}$ the rotational connection (see also~\cite{Edelen86}).
The torsion and the disclination density tensor are defined by
\begin{align}
\label{DislD2}
&\alpha_{ij}=\epsilon_{jkl}\pd_k\phi_{il} +\epsilon_{ikl} \Theta_{kj} x_l,\\
\label{DisclD2}
&\Theta_{ij}=\epsilon_{jmn}\pd_m W_{in}.
\end{align}
The disclination density tensor~(\ref{DisclD2}) is the linearized 
Riemann-Cartan curvature tensor or equivalent the corresponding Einstein tensor.
It can be seen that a non-vanishing  disclination tensor~(\ref{DisclD2}) 
gives a contribution to the torsion tensor~(\ref{DislD2}).
This piece may be called the disclination torsion. 
Therefore, the torsion~(\ref{DislD2}) has a contribution from both the 
translational sector (=dislocations) and the rotational sector (=disclinations).
Of course,
in the case of dislocations (teleparallelism), $\Theta_{ij}=0$, the disclination torsion is zero
and only the first piece in~(\ref{DislD2}), 
which is the proper dislocation density tensor, gives a non-vanishing
contribution.
If we compare Eq.~(\ref{DisclD1}) with~(\ref{DisclD2}),
we may identify (see also~\cite{Maly96})
\begin{align}
\label{W1}
W_{ij}\equiv-\varphi^*_{ij}.
\end{align}
Using Eqs.~(\ref{dist-GT}), (\ref{DisclD2}) and (\ref{W1}), one is
able to prove the equivalence between (\ref{DislD1}) and (\ref{DislD2}).

The basic equation for the force stress in an isotropic medium 
is the following inhomogeneous Helmholtz equation~\cite{Lazar02c}
\begin{align}
\label{stress-fe}
\Big(1-\kappa^{-2}\Delta\Big)\sigma_{ij}=\tl\sigma {}_{ij},\qquad \kappa^2=\frac{2\mu}{a_1},
\end{align}
where $\tl\sigma {}_{ij}$ is the stress tensor
obtained for the same traction boundary-value problem 
within the theory of classical elasticity.
It is important to note that (\ref{stress-fe}) agrees with the field equation
for the stress field in Eringen's nonlocal elasticity~\cite{Eringen83,Eringen85}
and in gradient elasticity~\cite{GA99}.
The factor $\kappa^{-1}$ has the physical dimension of a length and 
it defines, therefore, an internal characteristic length.
If we consider the two-dimensional problem and
using Green's function of the two-dimensional Helmholtz equation, 
we may solve the field equation for every component of the stress 
field~(\ref{stress-fe}) by the help of the convolution integral:
\begin{align}
\label{stress-nl}
\sigma_{ij}(r)=\int_V \alpha(r-r')\,\tl\sigma {}_{ij}(r')\, \d v(r'),
\end{align}
with the two-dimensional Green's function
\begin{align}
\label{green}
\alpha(r-r')& =\frac{\kappa^2}{2\pi}\,K_0(\kappa (r-r')),
\end{align}
with $r=\sqrt{x^2+y^2}$.
Here $K_n$ is the modified Bessel function of the second kind and 
$n=0,1,\ldots$ denotes the order of this function.
Thus,
\begin{align}
\Big(1-\kappa^{-2}\Delta\Big)\alpha(r)=\delta(r),
\end{align}
where $\delta(r):=\delta(x)\delta(y)$ denotes the two-dimensional
Dirac delta function.
In this way, we deduce Eringen's so-called nonlocal constitutive relation
for a linear homogeneous, isotropic solid with Green's function~(\ref{green}) 
as nonlocal kernel. This kernel~(\ref{green}) has its maximum at $r=r'$ and
describes the nonlocal interaction. 
Its two-dimensional volume-integral yields
\begin{align}
\int_V\alpha(r-r')\,\d v(r)=1,
\end{align}
and is the normalization condition of the nonlocal kernel.
In the classical limit ($\kappa^{-1}\rightarrow 0$), it becomes the Dirac delta function
\begin{align}
\lim_{\kappa^{-1}\to 0}\alpha(r-r')= \delta(r-r').
\end{align}
Note that Eringen~\cite{Eringen83,Eringen85,Eringen87,Eringen02} found the two-dimensional kernel~(\ref{green}) 
by giving the best match with the Born-K{\'a}rm{\'a}n model of the 
atomic lattice dynamics and the atomistic dispersion curves.
He used the choice $e_0=0.39$ for the length 
\begin{align}
\kappa^{-1}=e_0\,a,
\end{align}
where $a$ is an internal length (e.g. atomic lattice parameter)
and $e_0$ is a material constant. 

Using the inverse of the generalized Hooke's law~(\ref{Hooke-inv}) and (\ref{stress-fe}), we obtain an inhomogeneous
Helmholtz equation for every component of the strain tensor (see~\cite{Lazar02c})
\begin{align}
\label{strain-fe}
\Big(1-\kappa^{-2}\Delta\Big)E_{ij}=\tl E {}_{ij},
\end{align}
where $\tl E {}_{ij}$ is the classical strain tensor.
Equation~(\ref{strain-fe}) is similar to the equation for the strain in 
gradient theory used by Gutkin and Aifantis~\cite{GA96,GA97,GA99}
if we identify $\kappa^{-2}$ with the gradient coefficient
(see, e.g., equation~(4) in~\cite{GA99}).
Since the strain tensor fulfils an inhomogeneous Helmholtz equation,
we may rewrite~(\ref{strain-fe}) as a nonlocal relation for the strain
\begin{align}
\label{strain-nl}
E_{ij}(r)=\int_V \alpha(r-r')\,\tl E {}_{ij}(r')\, \d v(r'),
\end{align}
which is similar to the nonlocal relation for the stress~(\ref{stress-fe}).
Thus, field theory of elastoplasticity may be considered as a nonlocal
theory for the stress as well as the strain tensor. In contrast 
to Eringen's nonlocal theory where only the stress tensor has a nonlocal form.
We assume that the stress and strain fields at infinity should have the same form
for both the classical and elastoplastic field theory.

\section{Classical solution}
\setcounter{equation}{0}
In this section we present the ``classical'' stress field for a straight twist
disclination in an infinitely extended isotropic body by the help of the 
stress function method. 
We assume the disclination line is along the z-axis and the Frank vector has
the following form $\frank\equiv (0,\Omega,0)$.
In contrast to the case of a wedge disclination or screw and edge dislocations,
the situation is not really a two-dimensional problem for the twist disclination.
In the case of a straight twist disclination the three-dimensional space
may be considered as a product of the two-dimensional $xy$-plane and the 
independent one-dimensional $z$-line~\cite{deWit73b}. 
In this situation the $z$-axis plays a peculiar role.

The classical solution for the elastic stress fields was originally given by 
deWit~\cite{deWit73b}
\begin{align}
\label{t_xx}
&\tl\sigma {}_{xx}=-\frac{\mu \Omega}{2\pi(1-\nu)}\, 
\frac{zy\big(y^2+3x^2\big)}{r^4},\\
\label{t_yy}
&\tl\sigma {}_{yy}=-\frac{\mu \Omega}{2\pi(1-\nu)}\, 
\frac{zy\big(y^2-x^2\big) }{r^4},\\
\label{t_xy}
&\tl\sigma {}_{xy}=\frac{\mu \Omega}{2\pi(1-\nu)}\, 
\frac{zx\big(x^2-y^2\big)}{r^4},\\
\label{t_zz}
&\tl\sigma {}_{zz}=-\frac{\mu \Omega\nu }{\pi(1-\nu)}\, 
\frac{zy}{r^2},\\
\label{t_zx}
&\tl\sigma {}_{zx}=\frac{\mu \Omega}{2\pi(1-\nu)}\, 
\frac{xy}{r^2},\\
\label{t_zy}
&\tl\sigma {}_{zy}=-\frac{\mu \Omega}{2\pi(1-\nu)}\, 
\left\{(1-2\nu)\ln r+\frac{x^2}{r^2}\right\}.
\end{align}
Obviously, the expressions~(\ref{t_xx})--(\ref{t_zz}) contain
the classical singularity $\sim r^{-1}$ and 
a logarithmic singularity $\sim\ln r$ in~(\ref{t_zy}).
Thus, the classical elastic stress is infinite at the disclination line. 
The reason is that the classical theory of elasticity breaks down in 
the disclination core so that in the defect core region classical elasticity
fails to apply. Usually, the radius of this region is estimated by means
of atomic models.  
Due to the unphysical singularities it is erroneous to 
argue that the stress has a maximum/minimum value at the defect line.

For the situation of the strain condition, $\tl E {}_{zz}=0$,
Eqs.~(\ref{t_xx})--(\ref{t_zy}) can be calculated by using the so-called
stress function method in the following form
\begin{align}
\label{stress-ansatz-cl}
\tl\sigma {}_{ij}=
\left(\begin{array}{ccc}
\ \,  \pd^2_{yy}\tl f & -\pd^2_{xy}\tl f & -\pd_y \tl F\\
-\pd^2_{xy}\tl f & \pd^2_{xx}\tl f & \pd_x \tl F +\pd_z \tl g\\
-\pd_y \tl F& \pd_x\tl F+\pd_z\tl g&\tl p
\end{array}\right).
\end{align}
The stress is given in terms of the stress functions $\tl f$, $\tl F$, $\tl g$ and $\tl p$. 
In order to satisfy the force equilibrium the stress $\tl\sigma {}_{zz}$ 
has to fulfil the condition
\begin{align}
\tl p=\nu\Delta \tl f=-\pd_y \tl g,
\end{align}
where $\Delta\equiv\pd^2_{xx}+\pd^2_{yy}$ denotes the two-dimensional Laplacian.
The ``classical'' stress functions for the stress 
fields~(\ref{stress-ansatz-cl}) are
\begin{align}
\label{f_twist_cl}
&\tl f=-\frac{\mu \Omega}{2\pi(1-\nu)}\, z y \ln r,\\
\label{F_twist_cl}
&\tl F=-\frac{\mu \Omega}{2\pi(1-\nu)}\, x \ln r,\\ 
\label{g_twist_cl}
&\tl g=\frac{\mu \Omega\nu}{\pi(1-\nu)}\, z \ln r.
\end{align}
They satisfy the following two-dimensional differential equations
\begin{align}
\Delta\Delta\tl f&=-\frac{2\mu\Omega z}{(1-\nu)}\, \pd_y\delta(r),\\
\Delta\Delta\tl F&=-\frac{2\mu\Omega}{(1-\nu)}\, \pd_x\delta(r),\\
\Delta\tl g&=\frac{2\mu\Omega\nu z}{(1-\nu)}\, \delta(r).
\end{align}
Thus, $\tl f$ and $\tl F$ are biharmonic stress functions and 
$\tl g$ is a harmonic one. We see that $\tl F$ is
an Airy stress function, $\tl f$ is
an Airy stress function multiplied by $z$ and on the other hand
$\tl g$ is a Prandtl  stress function multiplied by $z$ (up to constant
pre-factors).

For convenience we give the classical elastic strain of the 
straight twist disclination (see~\cite{deWit73b})
\begin{align}
&\tl E {}_{xx}=-\frac{\Omega}{4\pi(1-\nu)}\, 
\frac{zy}{r^2}\bigg\{(1-2\nu)+\frac{2x^2}{r^2}\bigg\},\\
&\tl E {}_{yy}=-\frac{\Omega}{4\pi(1-\nu)}\, 
\frac{zy}{r^2}\bigg\{(1-2\nu)-\frac{2x^2}{r^2}\bigg\},\\
&\tl E {}_{xy}=\frac{\Omega}{4\pi(1-\nu)}\, 
\frac{zx}{r^2}\bigg\{1-\frac{2y^2}{r^2}\bigg\},\\
&\tl E {}_{zx}=\frac{\Omega}{4\pi(1-\nu)}\, 
\frac{xy}{r^2},\\
&\tl E {}_{zy}=-\frac{\Omega}{4\pi(1-\nu)}\, 
\bigg\{(1-2\nu)\ln r+\frac{x^2}{r^2}\bigg\},
\end{align}
which contains the ``classical'' singularities at $r=0$.

\section{Nonsingular solution}
\setcounter{equation}{0}
In this section we want to consider the twist disclination in the elastoplastic field theory 
to find modified solutions without the ``classical'' singularities.
The modified solutions are used to estimate the extent of disclination core,
thus providing information which cannot be obtained by using classical 
elasticity theory.

We make for the modified stress field an ansatz in terms 
of unknown stress functions which has the same form as the 
classical stress field~(\ref{stress-ansatz-cl})
\begin{align}
\label{stress-ansatz}
\sigma_{ij}=
\left(\begin{array}{ccc}
\ \,  \pd^2_{yy}f & -\pd^2_{xy}f & -\pd_y F\\
-\pd^2_{xy}f & \pd^2_{xx}f & \pd_x F +\pd_z g\\
-\pd_y F& \pd_x F+\pd_z g& p
\end{array}\right),
\end{align}
with the relation
\begin{align}
p=\nu\Delta f=-\pd_y g.
\end{align}
Substituting (\ref{stress-ansatz}) and (\ref{stress-ansatz-cl})
into (\ref{stress-fe}) we obtain three inhomogeneous Helmholtz equations
for the unknown stress functions
\begin{align}
\label{f_fe}
\Big(1-\kappa^{-2}\Delta\Big)f&=- \frac{\mu \Omega}{2\pi(1-\nu)}  \, z y\ln r ,\\
\label{F_fe}
\Big(1-\kappa^{-2}\Delta\Big)F&=- \frac{\mu \Omega}{2\pi(1-\nu)}\, x\ln r ,\\
\label{g_fe}
\Big(1-\kappa^{-2}\Delta\Big)g&=\frac{\mu \Omega\nu}{\pi(1-\nu)}\, z \ln r.
\end{align}
The inhomogeneous parts of~(\ref{f_fe})--(\ref{g_fe}) are the classical stress
functions. 
Using the same procedure as in the case of a straight edge dislocation (see~\cite{Lazar02c})
in order to solve the inhomogeneous Helmholtz equations, we can find the solutions
of ~(\ref{f_fe})--(\ref{g_fe}).
The solutions for the modified stress functions of a straight twist disclination 
are given by
\begin{align}
\label{f_twist}
f&=-\frac{\mu \Omega}{2\pi(1-\nu)}\, z y \bigg\{\ln r 
+\frac{2}{\kappa^2 r^2}\Big(1-\kappa r K_1(\kappa r)\Big)\bigg\},\\ 
\label{F_twist}
F&=-\frac{\mu \Omega}{2\pi(1-\nu)}\, x \bigg\{\ln r 
+\frac{2}{\kappa^2 r^2}\Big(1-\kappa r K_1(\kappa r)\Big)\bigg\},\\ 
\label{g_twist}
g&=\frac{\mu \Omega\nu}{\pi(1-\nu)}\, z\Big\{\ln r +K_0(\kappa r)\Big\},
\end{align}
where the first pieces are the classical stress 
functions~(\ref{f_twist_cl})--(\ref{g_twist_cl}).

\begin{figure}[tp]\unitlength1cm
\vspace*{-1.0cm}
\centerline{
(e)
\begin{picture}(8,6)
\put(0.0,0.2){\epsfig{file=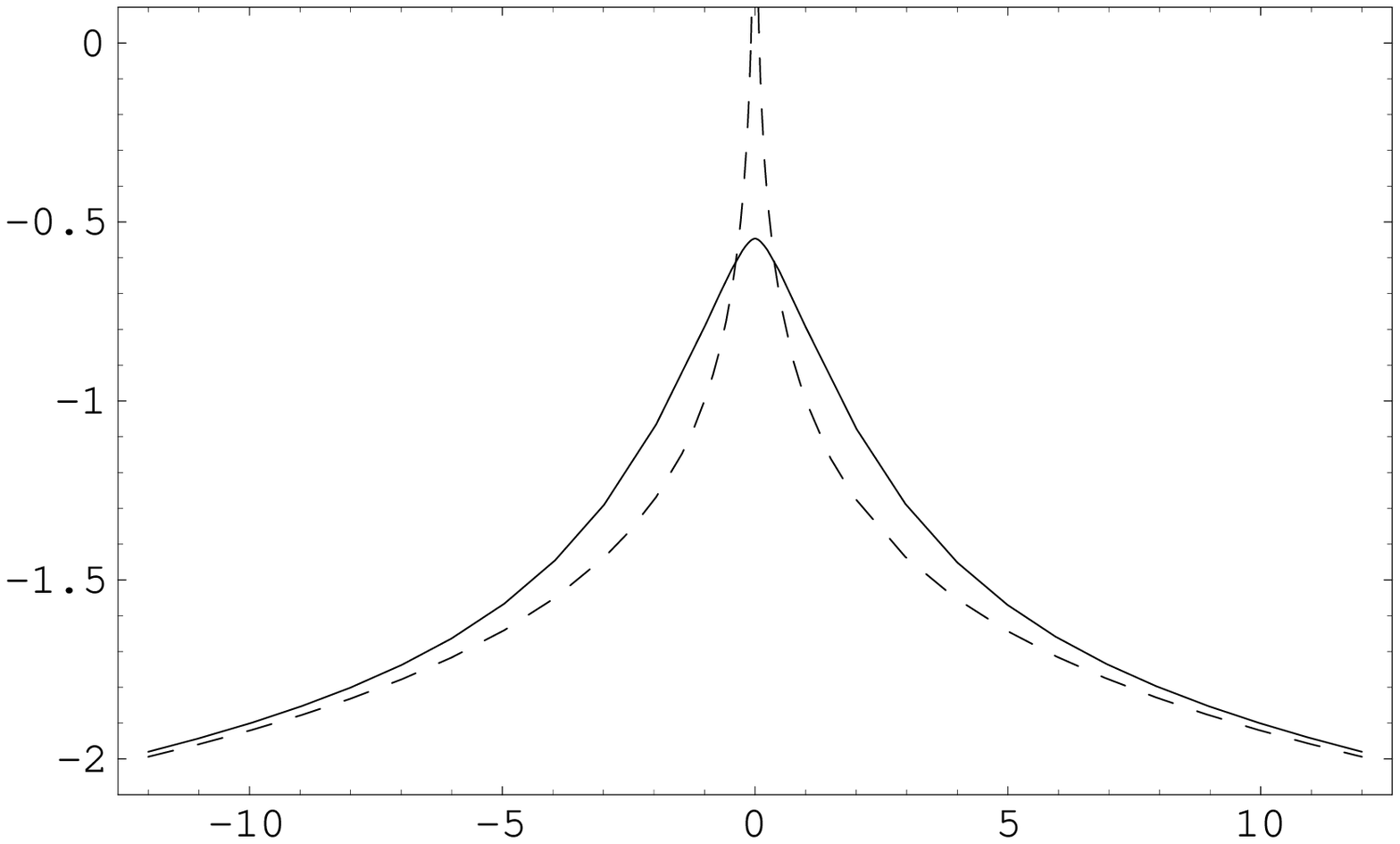,width=8cm}}
\put(4.0,0.0){$\kappa x$}
\put(-1.5,4.0){$\sigma_{zy}(x,0)$}
\end{picture}
}
\caption{The stress components of a twist disclination 
near the disclination line: 
(a) $\sigma_{xx}(0,y)$, (b) $\sigma_{yy}(0,y)$, 
(c) $\sigma_{xy}(x,0)$ are given in units of $\mu \Omega z\kappa/[2\pi(1-\nu)]$, 
(d) $\sigma_{zz}(0,y)$ is given in units of $\mu \Omega\nu \kappa/[\pi(1-\nu)]$
and 
(e) $\sigma_{zy}(x,0)$ is given in units of $\mu \Omega/[2\pi(1-\nu)]$.  
The dashed curves represent the classical stress components.}
\label{fig:stress2}
\end{figure}

By means of Eq.~(\ref{stress-ansatz}) and the stress 
functions~(\ref{f_twist})--(\ref{g_twist}),
we are able to calculate the modified stress of a straight twist disclination.
So we find for the elastic stress in Cartesian coordinates
\begin{align}
\label{T_xx}
&\sigma_{xx}=-\frac{\mu \Omega}{2\pi(1-\nu)}\, 
\frac{zy}{r^4}\bigg\{\big(y^2+3x^2\big)+\frac{4}{\kappa^2r^2}\big(y^2-3x^2\big)
-2 y^2\kappa r K_1(\kappa r)-2\big(y^2-3x^2\big) K_2(\kappa r)\bigg\},\\
\label{T_yy}
&\sigma_{yy}=-\frac{\mu \Omega}{2\pi(1-\nu)}\, 
\frac{zy}{r^4}\bigg\{\big(y^2-x^2\big)-\frac{4}{\kappa^2r^2}\big(y^2-3x^2\big)
-2 x^2\kappa r K_1(\kappa r)+2\big(y^2-3x^2\big) K_2(\kappa r)\bigg\},\\
\label{T_xy}
&\sigma_{xy}=\frac{\mu \Omega}{2\pi(1-\nu)}\, 
\frac{zx}{r^4}\bigg\{\big(x^2-y^2\big)-\frac{4}{\kappa^2r^2}\big(x^2-3y^2\big)
-2 y^2\kappa r K_1(\kappa r)+2\big(x^2-3y^2\big) K_2(\kappa r)\bigg\},\\
\label{T_zz}
&\sigma_{zz}=-\frac{\mu \Omega\nu }{\pi(1-\nu)}\, 
\frac{zy}{r^2}\Big\{1-\kappa r K_1(\kappa r)\Big\},\\
\label{T_zx}
&\sigma_{zx}=\frac{\mu \Omega}{2\pi(1-\nu)}\, 
\frac{xy}{r^2}\bigg\{1-\frac{2}{\kappa^2 r^2} \Big(2-\kappa^2 r^2 K_2(\kappa r)\Big)\bigg\},\\
\label{T_zy}
&\sigma_{zy}=-\frac{\mu \Omega}{2\pi(1-\nu)}\, 
\bigg\{(1-2\nu)\big(\ln r+K_0(\kappa r)\big)+\frac{x^2}{r^2}
-\frac{\big(x^2-y^2\big)}{\kappa^2 r^4} \Big(2-\kappa^2 r^2 K_2(\kappa r)\Big)\bigg\}.
\end{align}
These stresses are plotted in Fig.~\ref{fig:stress2}.
If we identify $\kappa\equiv 1/\sqrt{c}$ ($c$ is the gradient coefficient used by Gutkin and Aifantis),
the components of the stress~(\ref{T_xx})--(\ref{T_zy}) are in agreement with 
the stress field obtained by Gutkin and Aifantis~\cite{GA00,Gutkin00}
in the framework of strain gradient elasticity by using the Fourier transform
method.
It is interesting to note that the stresses~(\ref{T_xx})--(\ref{T_zz})
caused by the straight twist disclination with the Frank vector 
$\frank\equiv (0,\Omega,0)$ coincide with the stresses due to the
straight edge dislocation with the Burgers vector 
$\burger\equiv(b,0,0)$ replacing $\Omega z$ by $b$ (compare with
equations~(3.15)--(3.18) in~\cite{Lazar02c}).
The trace of the stress tensor $\sigma_{kk}=\sigma_{xx}+\sigma_{yy}+\sigma_{zz}$ 
produced by the twist disclination in an isotropic medium is
\begin{align}
\label{hyd_p}
\sigma_{kk}=-\frac{\mu \Omega(1+\nu)}{\pi(1-\nu)}\, 
\frac{zy}{r^2}\Big\{1-\kappa  r K_1(\kappa r)\Big\}.
\end{align}
We may now discuss some details of the stresses near the disclination core region
in the $xy$-plane.
The stresses~(\ref{T_xx})--(\ref{T_zz}) vanish at the disclination line instead
of being singular as predicted by classical elasticity.
Every component of~(\ref{T_xx})--(\ref{T_zz}) have a maximum and a minimum
near the disclination line. 
Because the extreme values are of opposite sign a zero point must be 
at the defect line.
In addition, the stress~(\ref{T_zy}) has a maximum value at the disclination line.
The extreme values may serve as a measure of critical stress level 
at which fracture or failure can occur. 
Contrary to classical elasticity, stresses~(\ref{T_xx})--(\ref{T_zy}) 
are finite at the defect line.
Therefore, the stress fields have no artificial singularities at the core  
and the maximum stress occurs at a short distance away from 
the disclination line (see~ Fig.~\ref{fig:stress2}). 
\begin{figure}[tp]\unitlength1cm
\vspace*{-1.0cm}
\centerline{
(a)
\begin{picture}(8,6)
\put(0.0,0.2){\epsfig{file=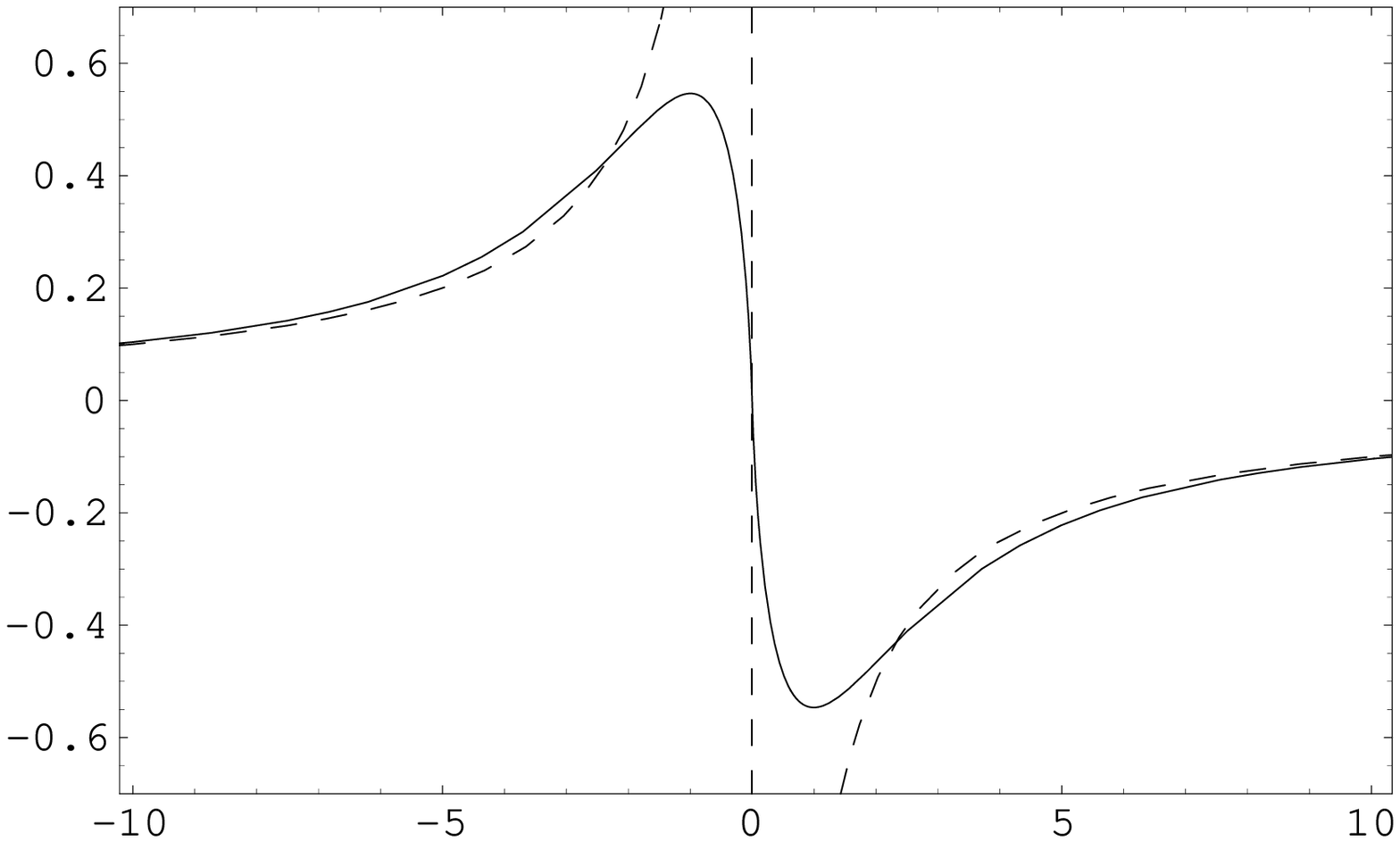,width=8cm}}
\put(4.0,0.0){$\kappa y$}
\put(-1.5,4.0){$\sigma_{xx}(0,y)$}
\end{picture}
}
\vspace*{-1.0cm}
\centerline{
(b)
\begin{picture}(8,6)
\put(0.0,0.2){\epsfig{file=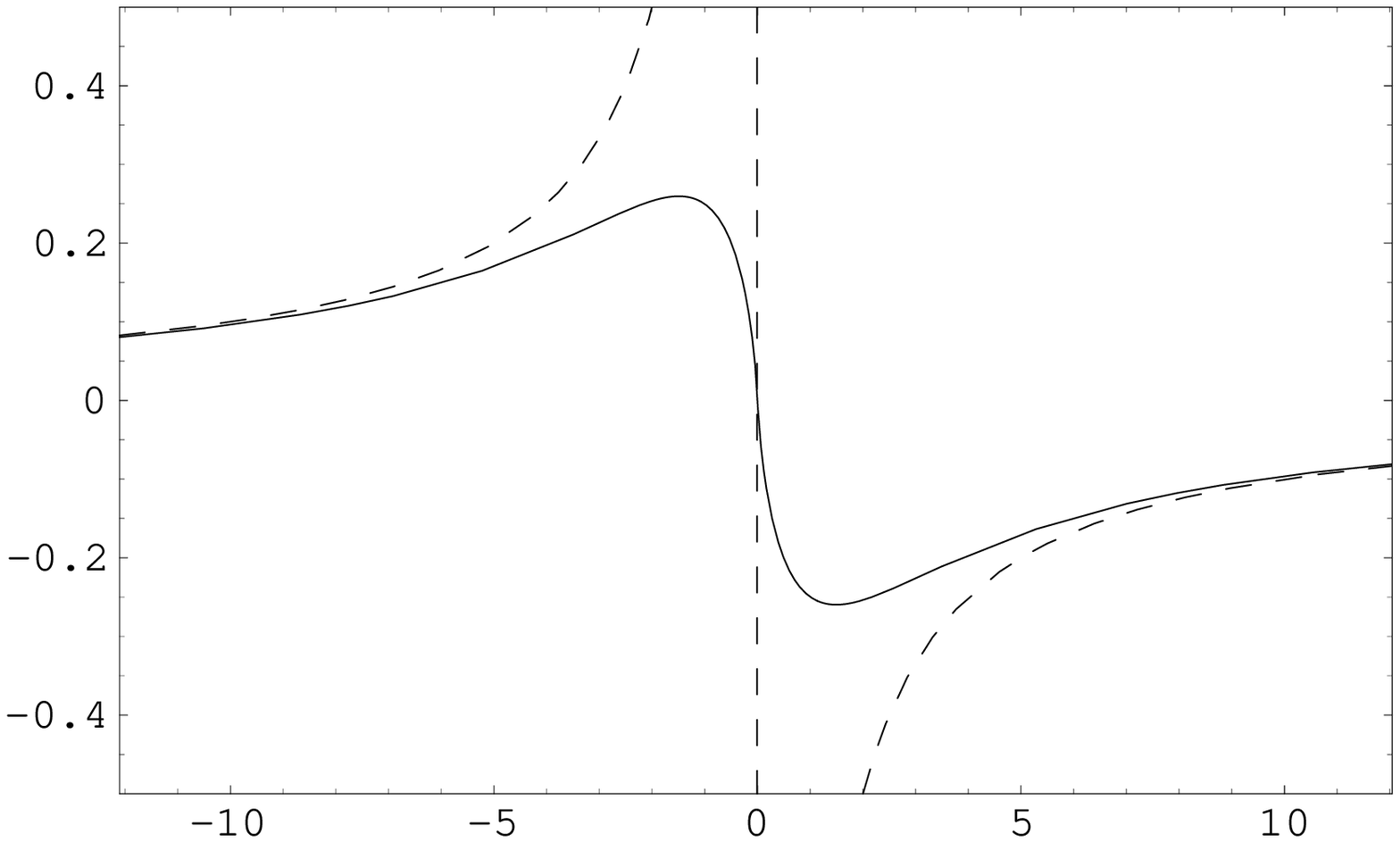,width=8cm}}
\put(4.0,0.0){$\kappa y$}
\put(-1.5,4.0){$\sigma_{yy}(0,y)$}
\end{picture}
}
\vspace*{-1.0cm}
\centerline{
(c)
\begin{picture}(8,6)
\put(0.0,0.2){\epsfig{file=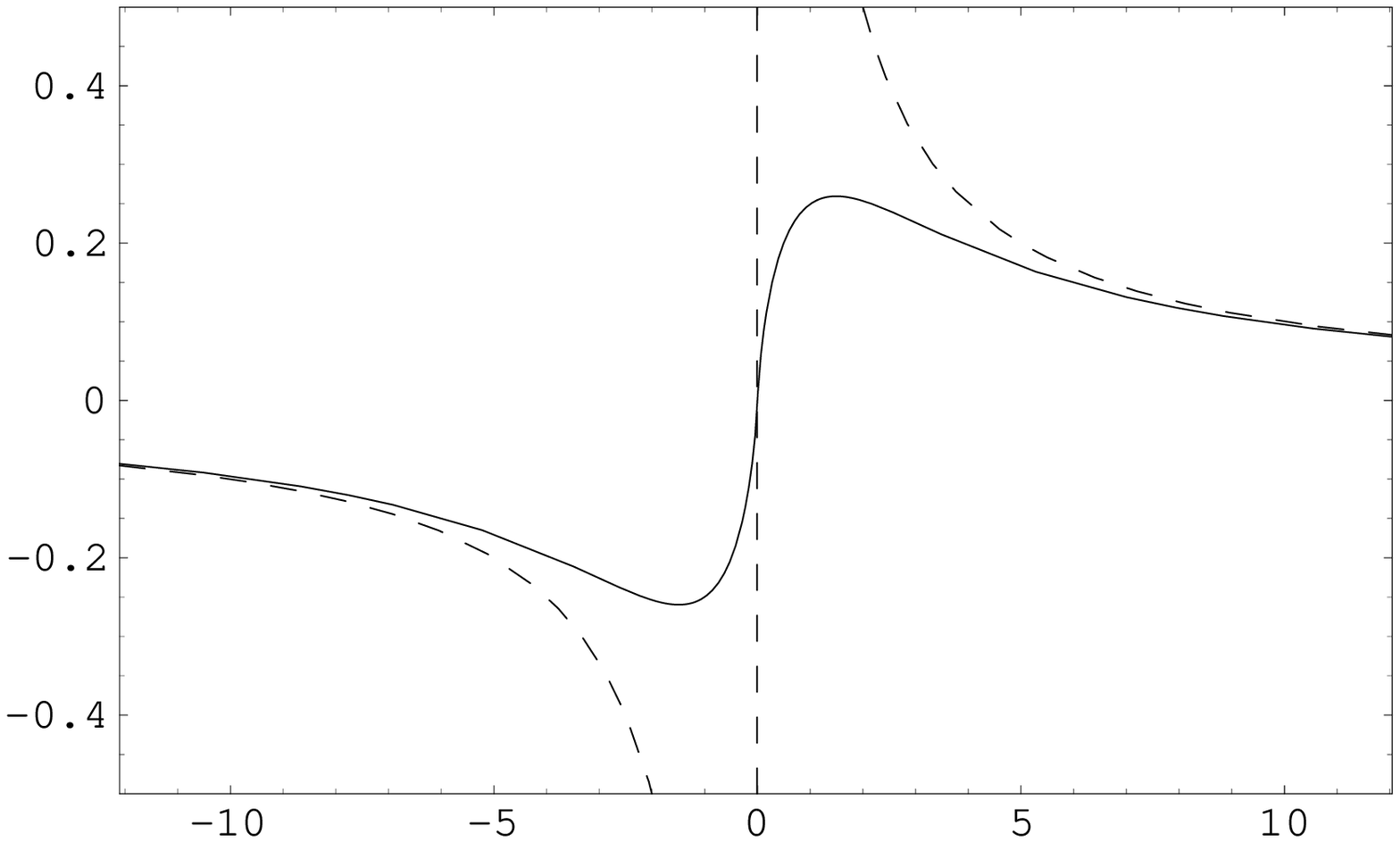,width=8cm}}
\put(4.0,0.0){$\kappa x$}
\put(-1.5,4.0){$\sigma_{xy}(x,0)$}
\end{picture}
}
\vspace*{-1.0cm}
\centerline{
(d)
\begin{picture}(8,6)
\put(0.0,0.2){\epsfig{file=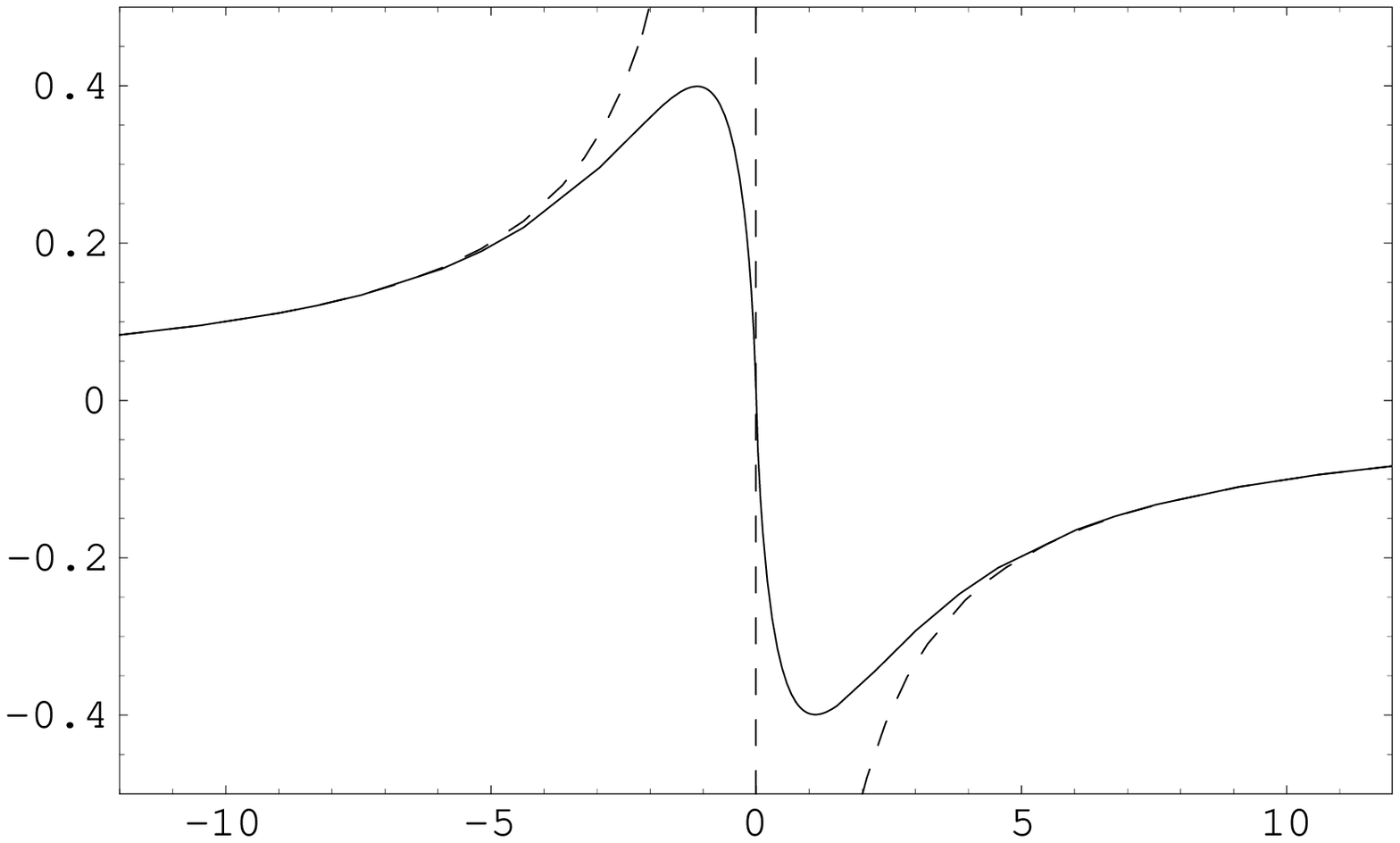,width=8cm}}
\put(4.0,0.0){$\kappa y$}
\put(-1.5,4.0){$\sigma_{zz}(0,y)$}
\end{picture}
}
\end{figure}
In fact, when $r\rightarrow 0$, we have
\begin{align*}
K_0(\kappa r)\rightarrow-\Big[\gamma+\ln \frac{\kappa r}{2}\Big],\qquad
K_1(\kappa r)\rightarrow \frac{1}{\kappa r},\qquad
K_2(\kappa r)\rightarrow -\frac{1}{2}+\frac{2}{(\kappa r)^2},
\end{align*}
and thus $\sigma_{ij}\rightarrow 0$. 
Here $\gamma$ denotes the Euler constant.
It can be seen that the stresses have the
following extreme values in the $xy$-plane:
$|\sigma_{xx}(0,y)|\simeq 0.546\kappa \frac{\mu \Omega z}{2\pi(1-\nu)}$ at 
$|y|\simeq 0.996 \kappa^{-1}$,
$|\sigma_{yy}(0,y)|\simeq 0.260 \kappa\frac{\mu \Omega z}{2\pi(1-\nu)}$ at 
$|y|\simeq 1.494 \kappa^{-1}$,
$|\sigma_{xy}(x,0)|\simeq 0.260 \kappa\frac{\mu \Omega z}{2\pi(1-\nu)}$ at 
$|x|\simeq 1.494 \kappa^{-1}$,
$|\sigma_{zz}(0,y)|\simeq 0.399\kappa \frac{\mu \Omega\nu z}{\pi(1-\nu)}$ at 
$|y|\simeq 1.114 \kappa^{-1}$ and
$|\sigma_{kk}(0,y)|\simeq 0.399\kappa \frac{\mu \Omega(1+\nu)z}{\pi(1-\nu)}$ at 
$|y|\simeq 1.114 \kappa^{-1}$.
The stresses $\sigma_{xx}$,  $\sigma_{yy}$ and  $\sigma_{xy}$ are modified 
near the disclination core ($0\le r\le 12\kappa^{-1}$). 
The stress $\sigma_{zz}$ and the trace $\sigma_{kk}$ are modified
in the region: $0\le r\le 6\kappa^{-1}$.
Far from the disclination line ($r\gg 12\kappa^{-1}$) the modified
and the classical solutions of the stress of a twist disclination coincide.
In addition, it can be seen that at $z=0$ the stresses~(\ref{T_xx})--(\ref{T_zz})
are zero. 
The stress $\sigma_{zy}$ has at $r=0$ the maximum value:
$\sigma_{zy}(0)\simeq \frac{\mu \Omega}{2\pi(1-\nu)}[(1-2\nu)(\gamma
+\ln\frac{\kappa}{2})-\frac{1}{2}]$ and with $\nu=0.3$:
$\sigma_{zy}(x,0)\simeq \frac{\mu \Omega}{2\pi(1-\nu)}[0.4\ln\kappa-0.546]$
(see Fig.~\ref{fig:stress2}e where a constant term
proportional to $\ln\kappa$ is dropped out).
Consequently, one can equate the maximum shear stresses to the cohesive shear stresses
to obtain conditions to produce a disclination of single atomic distance.

Due to the two-dimensional symmetry 
it is convenient to express the stresses in cylindrical coordinates.
The stress tensor has the following form in cylindrical coordinates
\begin{align}
\label{T_rr}
\sigma_{rr}&=-\frac{\mu \Omega}{2\pi(1-\nu)}\, 
\frac{z\sin\varphi}{r}\left\{1-\frac{4}{\kappa^2r^2}+2 K_2(\kappa r)\right\},\\
\label{T_rp}
\sigma_{r\varphi}&=\frac{\mu \Omega}{2\pi(1-\nu)}\, 
\frac{z\cos\varphi}{r}\left\{1-\frac{4}{\kappa^2r^2}+2 K_2(\kappa r)\right\},\\
\label{T_pp}
\sigma_{\varphi\varphi}&=-\frac{\mu \Omega}{2\pi(1-\nu)}\, 
\frac{z\sin\varphi}{r}\left\{1+\frac{4}{\kappa^2 r^2}
-2 K_2(\kappa r)-2\kappa r K_1(\kappa r)\right\},\\
\label{T_zz2}
\sigma_{zz}&=-\frac{\mu \Omega\nu }{\pi(1-\nu)}\, 
\frac{z\sin\varphi}{r}\Big\{1-\kappa r K_1(\kappa r)\Big\},\\
\label{T_zr}
\sigma_{zr}&=-\frac{\mu \Omega}{2\pi(1-\nu)}\, 
\sin\varphi
\bigg\{(1-2\nu)\big(\ln r+K_0(\kappa r)\big)
+\frac{2}{\kappa^2r^2}-K_2(\kappa r)\bigg\},\\
\label{T_zp}
\sigma_{z\varphi}&=-\frac{\mu \Omega}{2\pi(1-\nu)}\, 
\cos\varphi
\bigg\{(1-2\nu)\big(\ln r+K_0(\kappa r)\big)+1
-\frac{2}{\kappa^2r^2}+K_2(\kappa r)\bigg\}.
\end{align}
The fields~(\ref{T_rr})--(\ref{T_zp})
agree with the expressions given by Povstenko~\cite{Pov} 
in the framework of nonlocal elasticity if we use the identification
$\kappa\equiv 1/(\tau l)$. He used the two-dimensional nonlocal 
kernel~(\ref{green}) which is Green's function of the two-dimensional Helmholtz equation.
Again, if one replaces $\Omega z$ by $b$, the stresses~(\ref{T_rr})--(\ref{T_zp})
agree with the stresses of an edge dislocation in cylindrical coordinates
(compare with equations~(3.27)--(3.30) in~\cite{Lazar02c}).
The stresses~(\ref{T_rr})--(\ref{T_zp}) coincide with the classical ones 
far from the disclination core. 
The stresses~(\ref{T_rr})--(\ref{T_zz2}) are zero at the disclination line.
In principle we may discuss the extreme values of~(\ref{T_rr})--(\ref{T_zp}) 
in the $xy$-plane.
For example, the stresses~(\ref{T_rr}) and (\ref{T_pp}) have the extreme values: 
$|\sigma_{rr}|\simeq 0.260\kappa \frac{\mu \Omega z\sin\varphi}{2\pi(1-\nu)}$ at 
$r\simeq 1.494 \kappa^{-1}$ and
$|\sigma_{\varphi\varphi}|\simeq 0.547 \kappa\frac{\mu \Omega z\sin\varphi}{2\pi(1-\nu)}$ at 
$r\simeq 0.996 \kappa^{-1}$.
The stresses~(\ref{T_zr}) and (\ref{T_zp}) have the values at $r=0$:
$\sigma_{zr}(0)\simeq \frac{\mu \Omega \sin \varphi}{2\pi(1-\nu)}[(1-2\nu)(\gamma
+\ln\frac{\kappa}{2})-\frac{1}{2}]$ and  
$\sigma_{z\varphi}(0)\simeq \frac{\mu \Omega \cos \varphi}{2\pi(1-\nu)}[(1-2\nu)(\gamma
+\ln\frac{\kappa}{2})-\frac{1}{2}]$. 

The elastic strain is given in terms of stress functions 
\begin{align}
\label{stress-ansatz-strain}
E_{ij}=
\frac{1}{2\mu}\left(\begin{array}{ccc}
\pd^2_{yy}f -\nu\Delta f& -\pd^2_{xy}f & -\pd_y F\\
-\pd^2_{xy}f & \pd^2_{xx}f-\nu\Delta f& \pd_x F +\pd_z g\\
-\pd_y F& \pd_x F+\pd_z g& 0
\end{array}\right).
\end{align}
For the elastic strain of a twist disclination we find
\begin{align}
\label{E_xx}
&E_{xx}=-\frac{\Omega}{4\pi(1-\nu)}\, 
\frac{zy}{r^2}
\bigg\{(1-2\nu)+\frac{2x^2}{r^2}+\frac{4}{\kappa^2r^4}\big(y^2-3x^2\big)\\
&\hspace{5cm}-2\left(\frac{y^2}{r^2}-\nu\right)\kappa r K_1(\kappa r)
-\frac{2}{r^2}\big(y^2-3x^2\big) K_2(\kappa r)\bigg\},\nonumber\\
&E_{yy}=-\frac{\Omega}{4\pi(1-\nu)}\, 
\frac{zy}{r^2}
\bigg\{(1-2\nu)-\frac{2x^2}{r^2}-\frac{4}{\kappa^2r^4}\big(y^2-3x^2\big)\\
&\hspace{5cm}-2\left(\frac{x^2}{r^2}-\nu\right)\kappa r K_1(\kappa r)
+\frac{2}{r^2}\big(y^2-3x^2\big) K_2(\kappa r)\bigg\},\nonumber\\
&E_{xy}=\frac{\Omega}{4\pi(1-\nu)}\, 
\frac{zx}{r^2}
\bigg\{1-\frac{2y^2}{r^2}-\frac{4}{\kappa^2r^4}\big(x^2-3y^2\big)\\
&\hspace{5cm} -\frac{2y^2}{r^2}\,\kappa r K_1(\kappa r)
+\frac{2}{r^2}\big(x^2-3y^2\big) K_2(\kappa r)\bigg\},\nonumber\\
\label{E_zx}
&E_{zx}=\frac{\Omega}{4\pi(1-\nu)}\, 
\frac{xy}{r^2}\bigg\{1-\frac{2}{\kappa^2 r^2} \Big(2-\kappa^2 r^2 K_2(\kappa r)\Big)\bigg\},\\
\label{E_zy}
&E_{zy}=-\frac{\Omega}{4\pi(1-\nu)}\, 
\bigg\{(1-2\nu)\big(\ln r+K_0(\kappa r)\big)+\frac{x^2}{r^2}
-\frac{\big(x^2-y^2\big)}{\kappa^2 r^4} \Big(2-\kappa^2 r^2 K_2(\kappa r)\Big)\bigg\}.
\end{align}
The components of the strain tensor have in the $xy$-plane the
following extreme values ($\nu=0.3$):
$|E_{xx}(0,y)|\simeq 0.308\kappa \frac{\Omega z}{4\pi(1-\nu)}$ at 
$|y|\simeq 0.922 \kappa^{-1}$,
$|E_{yy}(0,y)|\simeq 0.010 \kappa\frac{\Omega z}{4\pi(1-\nu)}$ at 
$|y|\simeq 0.218 \kappa^{-1}$, 
$|E_{yy}(0,y)|\simeq 0.054 \kappa\frac{\Omega z}{4\pi(1-\nu)}$ at 
$|y|\simeq 4.130 \kappa^{-1}$, 
and
$|E_{xy}(x,0)|\simeq 0.260 \kappa\frac{\Omega z}{4\pi(1-\nu)}$ at 
$|x|\simeq 1.494 \kappa^{-1}$. 
It is interesting to note that 
$E_{yy}(0,y)$ is much smaller than $E_{xx}(0,y)$ within
the core region. 
The strain $E_{zy}$ has at $r=0$ the value:
$E_{zy}(0)\simeq \frac{\mu \Omega}{4\pi(1-\nu)}[(1-2\nu)(\gamma
+\ln\frac{\kappa}{2})-\frac{1}{2}]$.
The strain~(\ref{E_xx})--(\ref{E_zy}) coincides with 
the result given by Gutkin and Aifantis~\cite{GA00,Gutkin00,GA99b}.
The dilatation $E_{kk}$ reads
\begin{align}
E_{kk}=-\frac{\Omega(1-2\nu)}{2\pi(1-\nu)}\, 
\frac{zy}{r^2}\Big\{1-\kappa  r K_1(\kappa r)\Big\}.
\end{align}
In the $xy$-plane it has the extremum
$|E_{kk}(0,y)|\simeq 0.399\kappa \frac{\Omega(1-2\nu)z}{2\pi(1-\nu)}$ at 
$|y|\simeq 1.114 \kappa^{-1}$.
The elastic strain can be rewritten in cylindrical coordinates as follows
\begin{align}
\label{E_rr}
E_{rr}&=-\frac{\Omega}{4\pi(1-\nu)}\, 
\frac{z\sin\varphi}{r}\left\{(1-2\nu)-\frac{4}{\kappa^2r^2}+2 K_2(\kappa r)+2\nu K_1(\kappa r)\right\},\\
\label{E_rp}
E_{r\varphi}&=\frac{\Omega}{4\pi(1-\nu)}\, 
\frac{z\cos\varphi}{r}\left\{1-\frac{4}{\kappa^2r^2}+2 K_2(\kappa r)\right\},\\
\label{E_pp}
E_{\varphi\varphi}&=-\frac{\Omega}{4\pi(1-\nu)}\, 
\frac{z\sin\varphi}{r}\left\{(1-2\nu)+\frac{4}{\kappa^2 r^2}
-2 K_2(\kappa r)-2(1-\nu)\kappa r K_1(\kappa r)\right\},\\
\label{E_zr}
E_{zr}&=-\frac{\Omega}{4\pi(1-\nu)}\, 
\sin\varphi
\bigg\{(1-2\nu)\big(\ln r+K_0(\kappa r)\big)
+\frac{2}{\kappa^2r^2}-K_2(\kappa r)\bigg\},\\
\label{E_zp}
E_{z\varphi}&=-\frac{\Omega}{4\pi(1-\nu)}\, 
\cos\varphi
\bigg\{(1-2\nu)\big(\ln r+K_0(\kappa r)\big)+1
-\frac{2}{\kappa^2r^2}+K_2(\kappa r)\bigg\}.
\end{align}
The main feature of the solution given by~(\ref{E_xx})--(\ref{E_zp}) 
is the absence of any singularities near the disclination line.
This solution coincides with the classical ones far from the 
disclination core.

Now we want to calculate the elastic bend-twist, torsion, disclination density
and the rotation ($\omega_z\equiv-\beta_{[xy]}$ and $\omega_y\equiv\beta_{[xz]}$) 
of a twist disclination.
They might be determined from the following conditions on 
the dislocation densities of the twist disclination:
\begin{align}
\label{alpha_xz}
\alpha_{xz}&=-\frac{1-\nu}{2\mu}\,\pd_y\Delta f-\pd_x\omega_z,\\
\label{alpha_yz}
\alpha_{yz}&=\frac{1-\nu}{2\mu}\,\pd_x\Delta f-\pd_y\omega_z\equiv 0,\\
\label{alpha_zz}
\alpha_{zz}&=\frac{1}{2\mu}\big(\Delta F+\pd^2_{xz}g\big)-k_{zz}\equiv 0,\\
\label{alpha_xx}
\alpha_{xx}&=-\frac{1}{2\mu}\big(\pd^2_{yy} F
-\pd^3_{zxy} f\big)-k_{xx}\equiv 0,\\
\label{alpha_xy}
\alpha_{xy}&=\frac{1}{2\mu}\big(\pd^2_{xy} F
+\pd_z(\pd^2_{yy} f-\nu\Delta f)\big)-k_{yx}\equiv 0,\\
\label{alpha_yx}
\alpha_{yx}&=\frac{1}{2\mu}\big(\pd^2_{xy} F+\pd^2_{yz} g
-\pd_z(\pd^2_{xx} f-\nu\Delta f)\big)-k_{xy}\equiv 0,\\
\label{alpha_yy}
\alpha_{yy}&=-\frac{1}{2\mu}\big(\pd^2_{xx} F+\pd^2_{xz} g
+\pd^3_{zxy} f\big)-k_{yy}\equiv 0,\\
\label{alpha_zx}
\alpha_{zx}&=-\frac{1}{2\mu}\,\pd_z\big(\pd_{x} F+\pd_z g)-k_{xz}\equiv 0,\\
\label{alpha_zy}
\alpha_{zy}&=-\frac{1}{2\mu}\,\pd^2_{zy} F-k_{yz}\equiv 0,\\
\label{alpha-tr}
\alpha_{jj}&=2k_{jj}\equiv 0.
\end{align}
The conditions~(\ref{alpha_xz})--(\ref{alpha-tr}) must be consistent 
with deWit's dislocation densities of a twist disclination. 
We will discuss this point in detail below. 
The Eqs.~(\ref{alpha_xz}) and (\ref{alpha_yz}) look like the conditions
for the dislocation density of an edge dislocation (see~\cite{Lazar02c}) 
such that the elastic bend-twist $k_{zx}$ and $k_{zy}$ are compatible.
Eqs.~(\ref{alpha_zx}) and (\ref{alpha_zy}) are trivially satisfied.
From~(\ref{alpha_xz})--(\ref{alpha_yy}) we may determine the 
elastic bend-twist.
So we find for the non-vanishing components of the elastic bend-twist
tensor 
\begin{align}
\label{k_yx}
&k_{yx}=-\frac{\Omega}{2\pi}\,
\frac{y}{r^2}\Big\{1-\kappa r K_1(\kappa r)\Big\},\\
\label{k_yy}
&k_{yy}=\frac{\Omega}{2\pi}\,
\frac{x}{r^2}\Big\{1-\kappa r K_1(\kappa r)\Big\},\\
\label{k_zx}
&k_{zx}=\frac{\Omega z}{2\pi r^4}\Big\{\big(x^2-y^2\big)\big(1-\kappa r K_1(\kappa r)\big)
                -\kappa^2 x^2 r^2 K_0(\kappa r)\Big\}, \\
\label{k_zy}
&k_{zy}=\frac{\Omega z}{2\pi r^4}\, xy \Big\{2\big(1-\kappa r K_1(\kappa r)\big)
                -\kappa^2 r^2 K_0(\kappa r)\Big\},\\
\label{k_zz}
&k_{zz}=-\frac{\Omega}{2\pi}\,\frac{x}{r^2}\Big\{1-\kappa r K_1(\kappa r)\Big\}.
\end{align}
The shape of~(\ref{k_yx}) and (\ref{k_yy}) 
is analogous to the elastic bend-twist of a wedge disclination given 
in~\cite{Lazar03}.
It can be seen that (\ref{k_zx}) and (\ref{k_zy}) are singular at the 
disclination line $r=0$.
If one replaces $\Omega z$ by $b$, Eqs.~(\ref{k_yx}) and (\ref{k_yy}) 
coincide with the elastic bend-twist of an edge dislocation (see~\cite{Lazar02c,Lazar02d}).
The component~(\ref{k_zz}) has no singularity at the disclination line.

\begin{figure}[t]\unitlength1cm
\centerline{
(a)
\begin{picture}(9,6)
\put(0.0,0.2){\epsfig{file=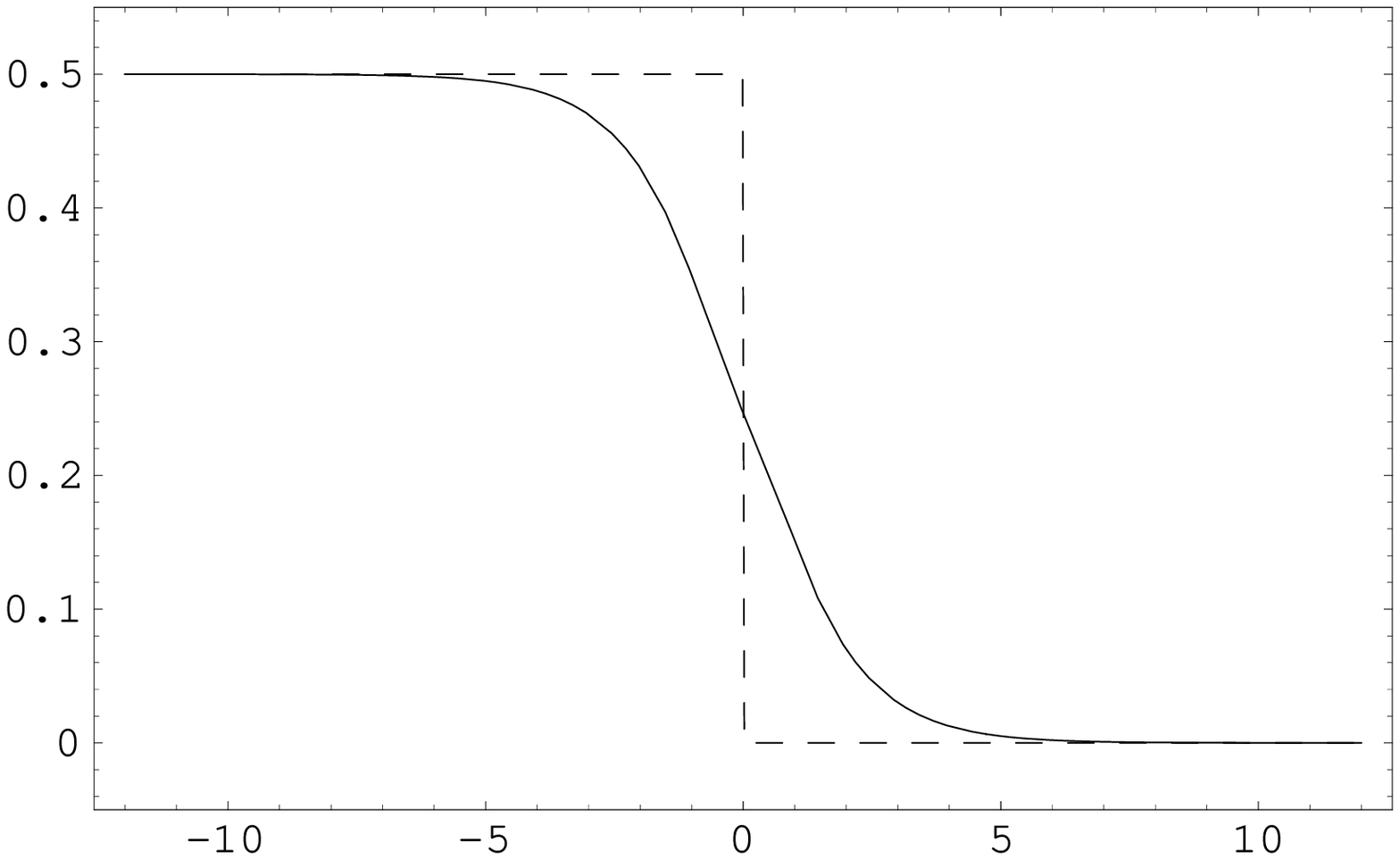,width=9cm}}
\put(4.5,0.0){$\kappa x$}
\put(-2.0,4.5){$\omega_y(x,0)/\Omega$}
\end{picture}
}
\centerline{
(b)
\begin{picture}(9,6)
\put(0.0,0.2){\epsfig{file=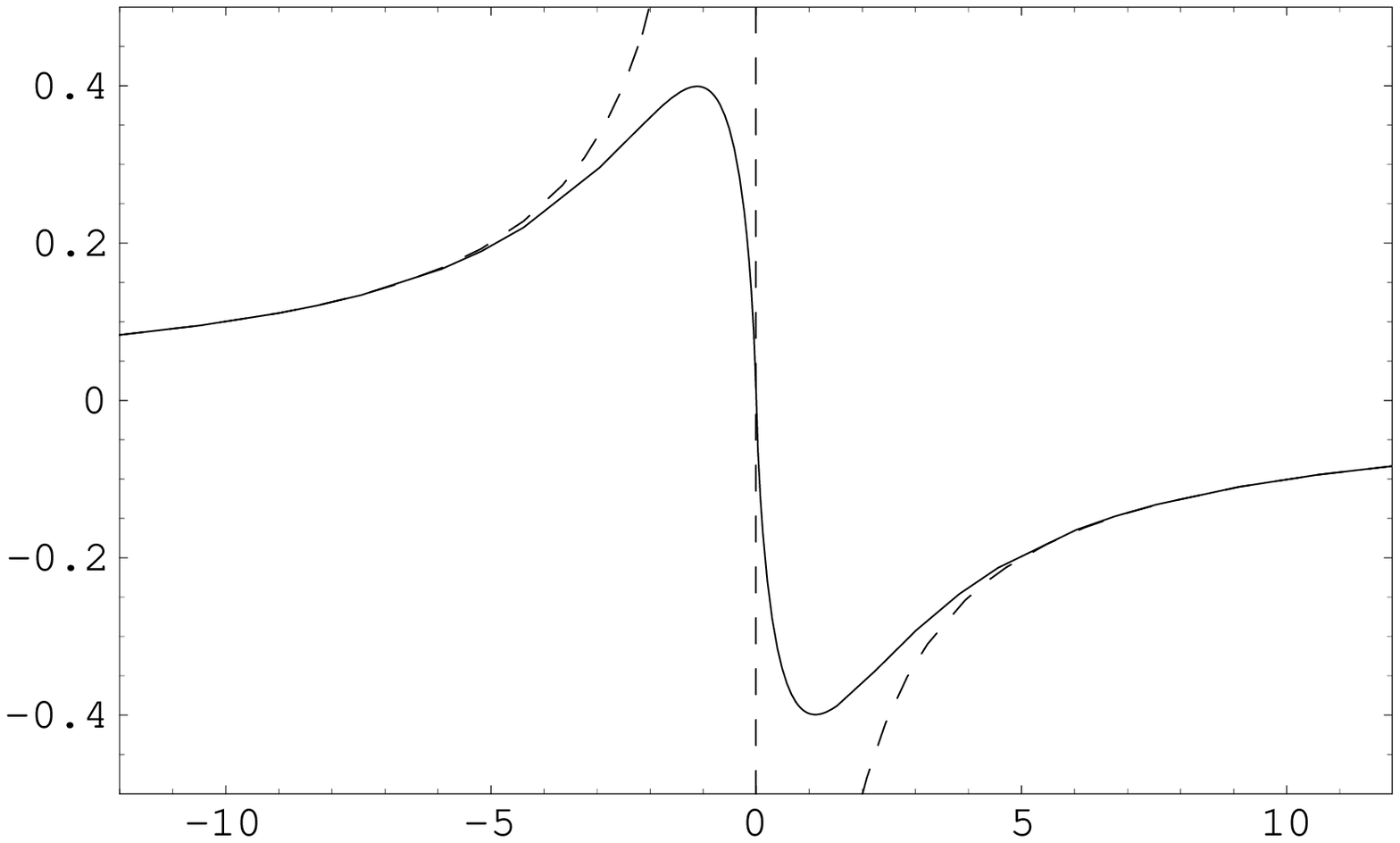,width=9.0cm}}
\put(4.5,0.0){$\kappa x$}
\put(-2.8,4.5){$\omega_z(x,0)\cdot 2\pi/[\Omega z \kappa]$}
\end{picture}
}
\caption{Rotation vector of a twist disclination:(a) $\omega_y(x,y\rightarrow+0)/\Omega$, 
(b) $\omega_z(x,0)$ is plotted in units of $\Omega z \kappa/[2\pi]$.
The dashed curves represent the classical solution.}
\label{fig:rot-vec}
\end{figure}

The elastic bend-twist tensor can be decomposed according to~(\ref{bt})
into a gradient of the rotation vector and an incompatible part.
We identify the incompatible part with the disclination loop density.
It is analogous to the decomposition of the elastic distortion of a dislocation
into a gradient of the displacement vector and an incompatible distortion 
(see~\cite{Lazar02a,Lazar02b,Lazar02c}).
We find for the non-vanishing components of the rotation vector
\begin{align}
\label{rot_y}
&\omega_y=
\frac{\Omega}{2\pi}\,\Big\{\varphi\big(1-\kappa r K_1(\kappa r)\big)
+\frac{\pi}{2}\,{\mathrm{sign}}(y)\,\kappa r K_1(\kappa r)\Big\},\\
\label{rot_z}
&\omega_z=-\frac{\Omega}{2\pi}\,\frac{zx}{r^2}\Big\{1-\kappa r K_1(\kappa r)\Big\}.
\end{align}
Here we use a single-valued discontinuous form for $\varphi$ 
(see~\cite{deWit73b,GA96,GA97,GA99}).
It is made unique by cutting the half-plane $y=0$ at $x<0$ and 
assuming $\varphi$ to jump from $\pi$ to $-\pi$ when crossing the cut.
The far fields of the rotation vector~(\ref{rot_y}) and (\ref{rot_z}) 
agree with deWit's expressions given in~\cite{deWit73b}. 
It yields ${\mathrm{sign}}\, y=+1$ for $y>0$ and 
${\mathrm{sign}}\, y=-1$ for $y<0$.
When $y\rightarrow +0$, the expression~(\ref{rot_y}) is plotted
in Fig.~\ref{fig:rot-vec}a. It can be seen that the Bessel function terms
which appear in~(\ref{rot_y}) lead to the symmetric smoothing
of the rotation vector profile, in contrast to the abrupt jump
occurring in the classical solution. It is interesting to note that the size
of such a transition zone is approximately $12/\kappa$ which gives 
the value $6/\kappa$ for the radius of the disclination core. 
The component~(\ref{rot_y}) is discontinuous due to $\varphi$ 
and (\ref{rot_z}) is continuous.
The component~(\ref{rot_z}) has in the $xy$-plane a maximum of
$\omega_z(x,0)\simeq 0.399\Omega\kappa z/[2\pi]$ at $x\simeq -1.114/\kappa$
and a minimum of
$\omega_z(x,0)\simeq -0.399\Omega\kappa z/[2\pi]$ at $x\simeq 1.114/\kappa$
and no singularity at the disclination core (see  Fig.~\ref{fig:rot-vec}b).
It can be seen that $k_{zx}$, $k_{zy}$ and $k_{zz}$ are gradient terms
of the rotation $\omega_z$.
In performing the differentiations of the rotation~$\omega_y$ 
we obtain $k_{yx}$ and $k_{yy}$ plus
excess terms which we identify as components of the 
disclination loop density.
The non-vanishing components of the disclination loop density turn out to be 
\begin{align}
\label{bt-pl}
&\varphi^*_{yx}=\frac{\Omega}{2\pi}\,
\kappa^2 x K_0(\kappa r)\Big(\varphi-\frac{\pi}{2}\,{\mathrm{sign}}(y)\Big),\\
&\varphi^*_{yy}=\frac{\Omega}{2\pi}\Big\{
\kappa^2 y K_0(\kappa r)\Big(\varphi
-\frac{\pi}{2}\,{\mathrm{sign}}(y)\Big)
+\pi\delta(y) \Big(1-{\mathrm{sign}}(x)\big[1-\kappa r K_1(\kappa r)\big]\Big)\Big\}.
\end{align}
They contain the angle $\varphi$ and the form is analogous to the plastic distortion 
of a dislocation~(see~\cite{Lazar02c}).
Only the component $\varphi^*_{yy}$ has a $\delta$-singularity 
at $y=0$ like the disclination loop density~\cite{Mura72,deWit73b} 
$\varphi^*_{yy}=(\Omega/2)\,\delta(y)(1-{\mathrm{sign}}(x))$.

Finally, we find for the elastic distortion of the straight twist disclination
\begin{align}
\label{dist_xx}
&\beta_{xx}=-\frac{\Omega}{4\pi(1-\nu)}\, 
\frac{zy}{r^2}
\bigg\{(1-2\nu)+\frac{2x^2}{r^2}+\frac{4}{\kappa^2r^4}\big(y^2-3x^2\big)\\
&\hspace{4cm}
-2\left(\frac{y^2}{r^2}-\nu\right)\kappa r K_1(\kappa r)
-\frac{2}{r^2}\big(y^2-3x^2\big) K_2(\kappa r)\bigg\},\nonumber\\
&\beta_{xy}=\frac{\Omega}{4\pi(1-\nu)}\, 
\frac{zx}{r^2}
\bigg\{(3-2\nu)-\frac{2y^2}{r^2}-\frac{4}{\kappa^2r^4}\big(x^2-3y^2\big)
\\
&\hspace{4cm} 
-2\left((1-\nu)+\frac{y^2}{r^2}\right)\kappa r K_1(\kappa r)
+\frac{2}{r^2}\big(x^2-3y^2\big) K_2(\kappa r)\bigg\},\nonumber\\
&\beta_{yx}=-\frac{\Omega}{4\pi(1-\nu)}\, 
\frac{zx}{r^2}
\bigg\{(1-2\nu)+\frac{2y^2}{r^2}+\frac{4}{\kappa^2r^4}\big(x^2-3y^2\big)\\
&\hspace{4cm}
-2\left((1-\nu)-\frac{y^2}{r^2}\right)\kappa r K_1(\kappa r)
-\frac{2}{r^2}\big(x^2-3y^2\big) K_2(\kappa r)\bigg\},\nonumber\\
\label{dist_yy}
&\beta_{yy}=-\frac{\Omega}{4\pi(1-\nu)}\, 
\frac{zy}{r^2}
\bigg\{(1-2\nu)-\frac{2x^2}{r^2}-\frac{4}{\kappa^2r^4}\big(y^2-3x^2\big)\\
&\hspace{4cm}
-2\left(\frac{x^2}{r^2}-\nu\right)\kappa r K_1(\kappa r)
+\frac{2}{r^2}\big(y^2-3x^2\big) K_2(\kappa r)\bigg\}\nonumber\\
\label{dist_zx}
&\beta_{zx}=\frac{\Omega}{4\pi(1-\nu)}\, 
\frac{xy}{r^2}\bigg\{1-\frac{2}{\kappa^2 r^2} \Big(2-\kappa^2 r^2 K_2(\kappa r)\Big)\bigg\}\\
&\hspace{4cm}
-\frac{\Omega}{2\pi}\,
\Big\{\varphi\big(1-\kappa r K_1(\kappa r)\big)
+\frac{\pi}{2}\,{\mathrm{sign}}(y)\,\kappa r K_1(\kappa r)\Big\},\nonumber\\
\label{dist_xz}
&\beta_{xz}=\frac{\Omega}{4\pi(1-\nu)}\, 
\frac{xy}{r^2}\bigg\{1-\frac{2}{\kappa^2 r^2} \Big(2-\kappa^2 r^2 K_2(\kappa r)\Big)\bigg\}\\
&\hspace{4cm}
+\frac{\Omega}{2\pi}\,
\Big\{\varphi\big(1-\kappa r K_1(\kappa r)\big)
+\frac{\pi}{2}\,{\mathrm{sign}}(y)\,\kappa r K_1(\kappa r)\Big\},\nonumber\\
\label{dist_zy}
&\beta_{zy}=-\frac{\Omega}{4\pi(1-\nu)}
\bigg\{ (1-2\nu)\big(\ln r +K_0(\kappa r)\big)
+\frac{x^2}{r^2}-\frac{(x^2-y^2)}{\kappa^2 r^4}\Big(2-\kappa^2 r^2 K_2(\kappa r)\Big)\bigg\},\\
\label{dist_yz}
&\beta_{yz}=-\frac{\Omega}{4\pi(1-\nu)}
\bigg\{ (1-2\nu)\big(\ln r +K_0(\kappa r)\big)
+\frac{x^2}{r^2}-\frac{(x^2-y^2)}{\kappa^2 r^4}\Big(2-\kappa^2 r^2 K_2(\kappa r)\Big)\bigg\}.
\end{align}
Replacing $\Omega z$ by $b$, Eqs.~(\ref{dist_xx})--(\ref{dist_yy}) are analogous 
to the elastic distortion of an edge dislocation (see~\cite{Lazar02c}).
The components of the elastic distortion~(\ref{dist_zx}) and (\ref{dist_xz}) 
contain the angle $\varphi$ in contrast to the dislocation case. 
But this is a typical property of a disclination.

\begin{figure}[t]\unitlength1cm
\centerline{
\begin{picture}(9,6)
\put(0.0,0.2){\epsfig{file=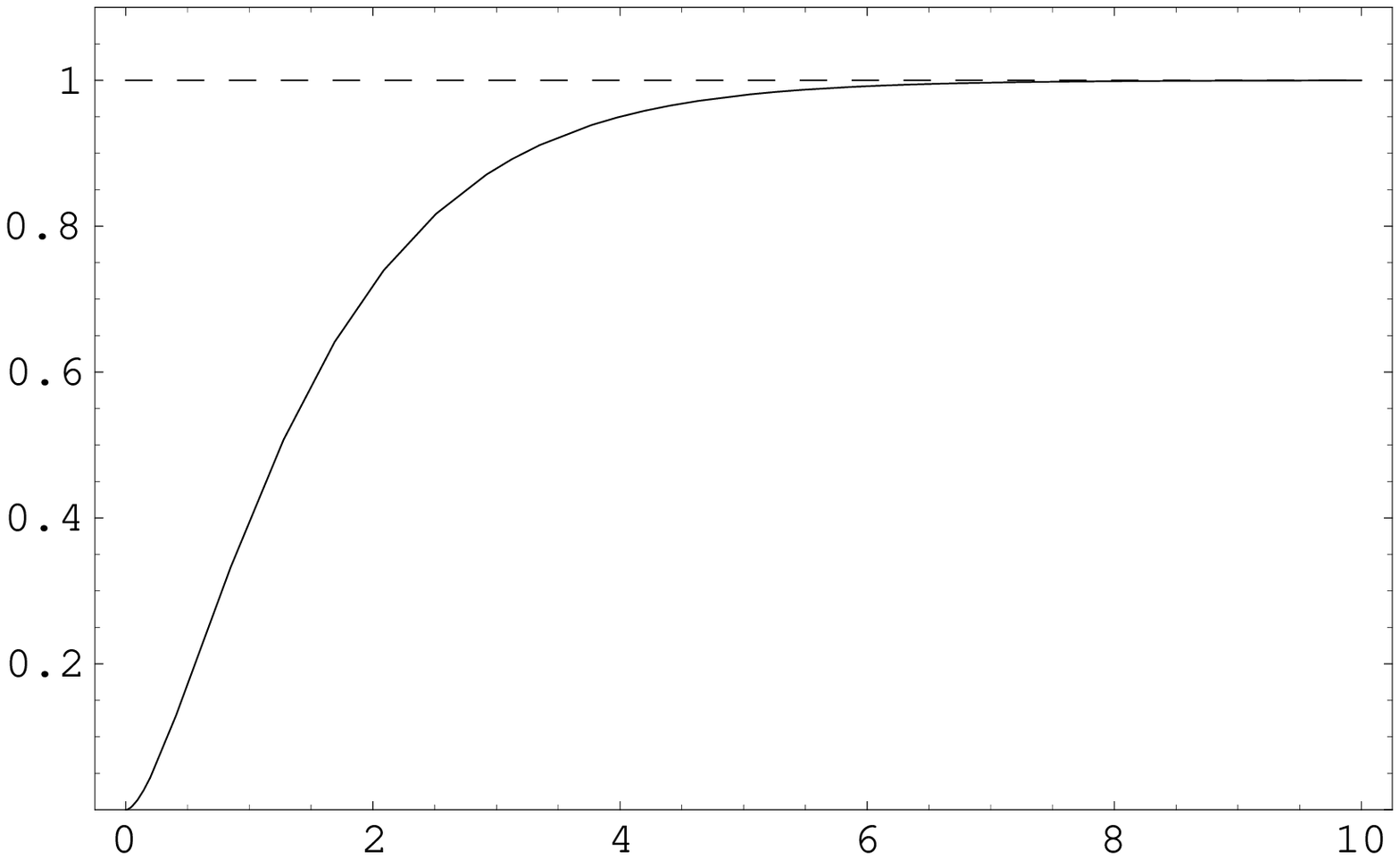,width=9cm}}
\put(4.5,0.0){$\kappa r$}
\put(-1.7,4.5){$\Omega_y(r)/\Omega$}
\end{picture}
}
\caption{Effective Frank vector $\Omega_y(r)/\Omega$ (solid).}
\label{fig:frank}
\end{figure}

With Eq.~(\ref{Frank})
we obtain for the effective Frank vector of the twist disclination
\begin{align}
\label{Frank-eff}
\Omega_y(r)=\oint_\gamma \big(k_{yx}\d x+k_{yy}\d y\big)
        =\Omega\,\Big\{1-\kappa r K_1(\kappa r)\Big\}.
\end{align}
It differs appreciably from the constant value $\Omega$ in the 
region from $r=0$ up to $r\simeq 6/\kappa$ (see Fig.~\ref{fig:frank}). 
In fact, we find $\Omega_y(0) =0$ and $\Omega_y(\infty)=\Omega$.
Thus, it is suggestive to
take $r_c\simeq 6/\kappa$ as the core radius of the disclination.
The effective Frank vector $\Omega_y(r)$ of a straight twist disclination
has the same form as the effective Frank vector $\Omega_z(r)$ of 
a straight wedge disclination which is given in~\cite{Lazar03}.

In the case of a twist disclination  
we obtain the following disclination torsion
\begin{align}
\label{dd-xz}
\alpha_{xz}=\frac{\Omega\kappa^2}{2\pi}\,   z\, K_0(\kappa r).
\end{align}
It looks like a dislocation density of a straight edge dislocation whose ``Burgers vector'' $\Omega z$ 
depends on the position $z$. 
At the point $z=0$ the dislocation density~(\ref{dd-xz}) is zero.
The dislocation line of the edge dislocation coincides 
with the disclination line of the twist disclination.
Therefore, this dislocation density implies a dislocation line
with changing Burgers vector in agreement with deWit~\cite{deWit73b}. 
In the limit $1/\kappa\rightarrow 0$, deWit's classical 
expression $\alpha_{xz}=\Omega z\, \delta(r)$ is restored. 
Consequently, we found that the straight twist disclination contains 
a certain amount of dislocation density (see also~\cite{deWit73b}).

The elastic distortion gives rise to an effective Burgers vector
\begin{align}
\label{Burgers-eff}
b_x(r)=\oint_\gamma \big(\beta_{xx}\d x+\beta_{xy}\d y\big)
        =\Omega z\,\Big\{1-\kappa r K_1(\kappa r)\Big\}.
\end{align}
We see explicitly the changing of the Burgers vector on the dislocation
and disclination line. 
The effective Burgers vector differs from the constant value $\Omega z$ in the 
region from $r=0$ to $r\simeq 6/\kappa$. 
We find $b_x(0)=0$ and $b_x(\infty)=\Omega z$.
In addition, the Burgers vector~(\ref{Burgers-eff}) depends on the position of $z$. At the position $z=0$
it is zero.
From~(\ref{Frank-eff}) and (\ref{Burgers-eff}) we obtain the relation between
the effective Burgers and Frank vector
\begin{align}
b_x(r)=z\, \Omega_y(r).
\end{align}

We find for the non-vanishing component of the disclination density~(\ref{DisclD1}) 
of a twist disclination
\begin{align}
\label{den-twist}
\Theta_{yz}=\frac{\Omega\kappa^2}{2\pi}\, K_0(\kappa r).
\end{align}
In the limit as $\kappa^{-1}\rightarrow0$, 
the result~(\ref{den-twist}) converts to the classical expression
$\Theta_{yz}=\Omega\, \delta(r)$.
The disclination density tensor~(\ref{den-twist}) and the disclination torsion~(\ref{dd-xz}) 
of a twist disclination fulfil the Eq.~(\ref{DislD2}) as follows
\begin{align}
\alpha_{xz}=z\, \Theta_{yz}.
\end{align}
Since the dislocation density~(\ref{dd-xz}) and the disclination density~(\ref{den-twist}) 
are localized at the same position it seems that the dislocation density (disclination torsion) 
is coupled on the twist disclination. It is a characteristic quantity of a 
twist disclination which 
cannot be created by a pure edge dislocation without the
presence of a twist disclination. In general, the disclination torsion is not
independent of the disclination density of a straight twist disclination. 
Only in the $xy$-plane at $z=0$ 
the disclination torsion and the Burgers vector of the corresponding 
twist disclination are zero.

\begin{figure}[p]\unitlength1cm
\centerline{
(a)
\begin{picture}(9,6)
\put(0.0,0.2){\epsfig{file=wy.eps,width=9cm}}
\put(4.5,0.0){$\kappa x$}
\put(-2.5,4.5){$u_x(x,0)/[\Omega z]$}
\end{picture}
}
\centerline{
(b)
\begin{picture}(9,6)
\put(-0.4,0.2){\epsfig{file=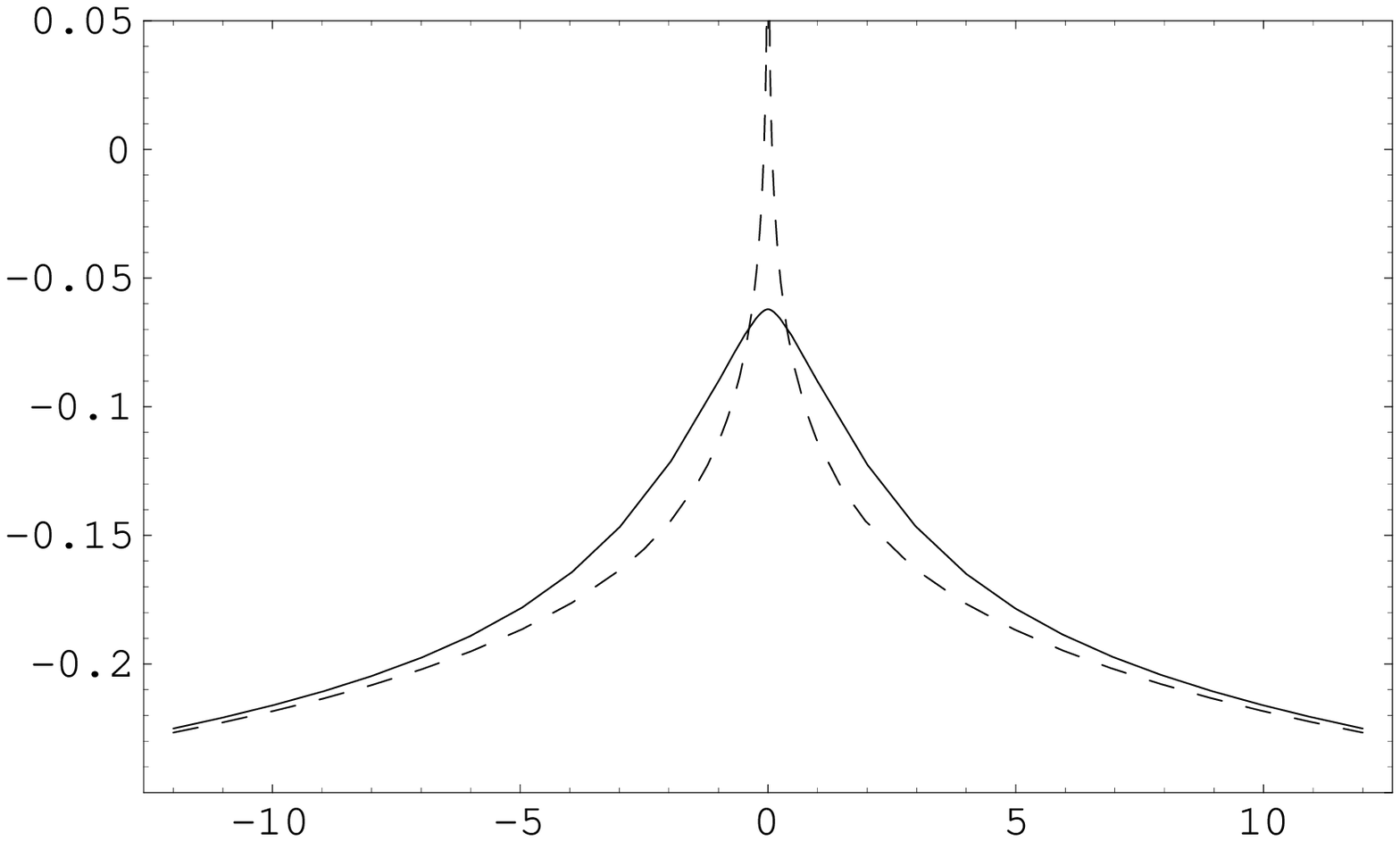,width=9.4cm}}
\put(4.5,0.0){$\kappa x$}
\put(-2.5,4.5){$u_y(x,0)/[\Omega z]$}
\end{picture}
}
\centerline{
(c)
\begin{picture}(9,6)
\put(0.3,0.2){\epsfig{file=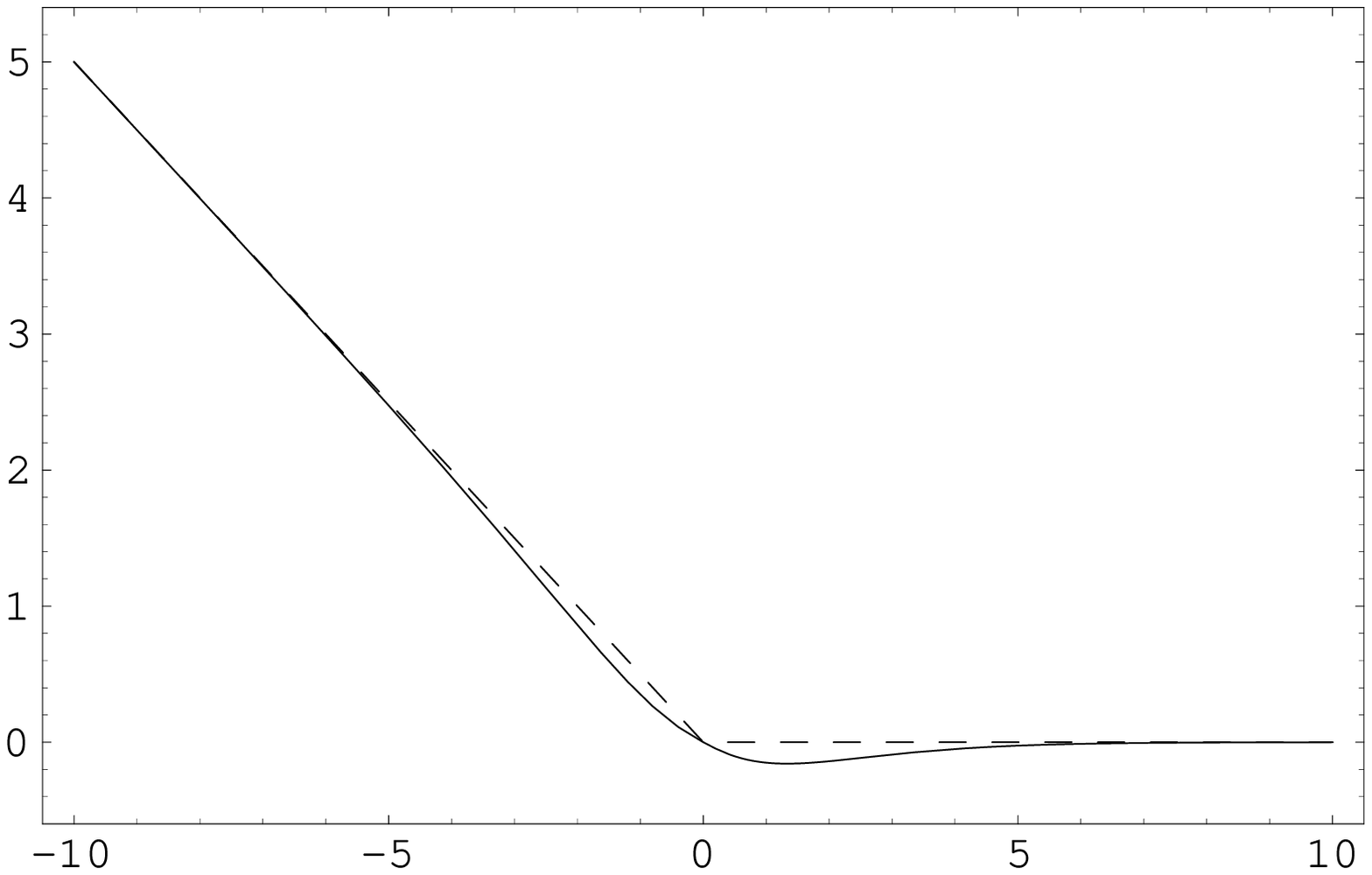,width=8.7cm}}
\put(4.5,0.0){$\kappa x$}
\put(-2.5,4.5){$u_z(x,0)/\Omega$}
\end{picture}
}
\caption{Displacement vector of a twist disclination: 
(a) $u_x(x,y\rightarrow+0)/[\Omega z]$, 
(b) $u_y(x,0)/[\Omega z]$ with $\nu=0.3$
(c) $u_z(x,y\rightarrow+0)/\Omega$.
The dashed curves represent the classical solution.}
\label{fig:u-vec}
\end{figure}

If we use the decomposition~(\ref{dist2}) for the distortions~(\ref{dist_xx})--(\ref{dist_xz}),
we may restore an effective displacement field and a properly incompatible
distortion. The displacement is given by
\begin{align}
\label{u_x}
u_x&=\frac{\Omega z}{2\pi}\bigg\{\varphi\big(1-\kappa r K_1(\kappa r)\big)
+\frac{\pi}{2}\, {\mathrm{sign}}(y)\,\kappa r K_1(\kappa r)
+\frac{1}{2(1-\nu)}\, \frac{xy}{r^2}\left(1-\frac{4}{\kappa^2r^2}+2 K_2(\kappa r)\right)\bigg\},\\
\label{u_y}
u_y&=-\frac{\Omega z}{4\pi(1-\nu)}\bigg\{(1-2\nu)\big(\ln r+ K_0(\kappa r)\big)
+\frac{x^2}{r^2}-\frac{(x^2-y^2)}{\kappa^2 r^4}\,\big(2-\kappa^2 r^2 K_2(\kappa r)\big)\bigg\},\\
\label{u_z}
u_z&=-\frac{\Omega}{2\pi}\bigg\{
        x\Big[\varphi\big(1-\kappa r K_1(\kappa r)\big)
        +\frac{\pi}{2}\,{\mathrm{sign}}(y)\,\kappa r K_1(\kappa r)\Big]\nonumber\\
        &\qquad\qquad
        +\frac{y}{2(1-\nu)}\Big[(1-2\nu)\big(\ln r -1 +K_0(\kappa r)\big)
        -\frac{1}{\kappa^2 r^2}\,\big(2-\kappa^2 r^2 K_2(\kappa r)\big)\Big]\bigg\}.
\end{align}
These displacements~(\ref{u_x})--(\ref{u_z}) have no singularities at the
disclination line.
When $y\rightarrow 0$, the Bessel function terms in ~(\ref{u_x})
lead to the symmetric smoothing of the displacement
profile in contrast to the abrupt jump occurring in the profile of the classical
solution (see Fig.~\ref{fig:u-vec}a).
Eqs.~(\ref{u_y}) and (\ref{u_z})
demonstrate the elimination of ``classical'' logarithmic singularities 
at the disclination line (see Fig.~\ref{fig:u-vec}b where a constant term
proportional to $\ln\kappa$ is neglected). 
When $y\rightarrow 0$, the Bessel function terms in ~(\ref{u_z})
smooth the displacement profile in the core region (see Fig.~\ref{fig:u-vec}c).
It is interesting to note that the non-classical parts of the displacements~(\ref{u_x}) and (\ref{u_y})
caused by the straight twist disclination with the Frank vector 
$\frank\equiv (0,\Omega,0)$ coincide with the non-classical parts of the 
displacements due to the straight edge dislocation with the Burgers vector 
$\burger\equiv(b,0,0)$ if we replace $\Omega z$ by $b$ (compare with
equations~(3.47) and (3.52) in~\cite{Lazar02c}). In addition, 
the displacement~(\ref{u_z}) coincides with the displacement $-u_y$ of
a wedge disclination (compare equation~(43) for $C=0$ in~\cite{Lazar03}).
The classical parts of (\ref{u_x})--(\ref{u_z}) agree with the displacement
given by deWit~\cite{deWit73b}.
The displacement values should be detectable for nanoparticle 
containing twist disclinations. Therefore, one could compare the 
displacements~(\ref{u_x})--(\ref{u_z}) with experimental and simulated 
results.
In performing the differentiations of the displacement~(\ref{u_x})--(\ref{u_z}) 
we obtain the total distortion $\beta^T_{ij}\equiv\pd_j u_i$.
Using~(\ref{dist2}) and
comparing the total distortion with the elastic one~(\ref{dist_xx})--(\ref{dist_yz}) 
the excess terms of the total distortion may be identified with the plastic
part.
So the incompatible distortion can be found as
\begin{align}
\label{plastic-dist-xx}
&\tilde\beta_{xx}=-\frac{\Omega z}{2\pi}\,\kappa^2 x K_0(\kappa r)
\left(\varphi-\frac{\pi}{2}\, {\mathrm{sign}}(y)\right),\\
&\tilde\beta_{xy}=-\frac{\Omega z}{2\pi}
\label{plastic-dist-xy}
\left\{\kappa^2 y K_0(\kappa r)\left(\varphi-\frac{\pi}{2}\, {\mathrm{sign}}(y)\right)
+\pi\delta(y) \Big(1-{\mathrm{sign}}(x)\big[1-\kappa r K_1(\kappa r)\big]\Big)\right\},\\
\label{plastic-dist-zx}
&\tilde\beta_{zx}=\frac{\Omega}{2\pi}\, \kappa^2 x^2 K_0(\kappa r)
\Big(\varphi-\frac{\pi}{2}\,{\mathrm{sign}}(y)\Big),\\
\label{plastic-dist-zy}
&\tilde\beta_{zy}=-\frac{\Omega}{2\pi}\,\Big\{\kappa r K_1(\kappa r)
-\kappa^2 x y K_0(\kappa r)
\Big(\varphi-\frac{\pi}{2}\,{\mathrm{sign}}(y)\Big)
-\pi\delta(y)\, x \Big(1-{\mathrm{sign}}(x)\big[1-\kappa r K_1(\kappa r)\big]\Big)\Big\}.
\end{align}
Eqs.~(\ref{plastic-dist-xx})--(\ref{plastic-dist-zy}) satisfy the relation~(\ref{tors1}).
The $\delta$-terms in~(\ref{plastic-dist-xy}) and (\ref{plastic-dist-zy}) 
have a similar form like deWit's
plastic distortion~\cite{deWit73b} of a twist disclination
$\beta^P_{xy}=(\Omega z/2)\,\delta(y)(1-{\mathrm{sign}}(x))$ and
$\beta^P_{zy}=-(\Omega x/2)\,\delta(y)(1-{\mathrm{sign}}(x))$.
But now the singularity surface is not strictly bonded by the 
disclination line. 
The incompatible distortions~(\ref{plastic-dist-xx}) and (\ref{plastic-dist-xy})
coincide with the incompatible distortion of an edge dislocation 
if we replace $\Omega z$ by $b$ (see~\cite{Lazar02c})
and (\ref{plastic-dist-zx}) and (\ref{plastic-dist-zy}) 
agree with the incompatible distortion $-\tilde\beta_{yx}$ and $-\tilde\beta_{yy}$ of a wedge 
disclination (see~\cite{Lazar03}).

Using~(\ref{dist-GT}) we obtain for~(\ref{plastic-dist-xx})--(\ref{plastic-dist-zy}) 
the following decomposition
\begin{align}
&\tilde\beta_{xx}=z W_{yx},\nonumber\\
&\tilde\beta_{xy}=z W_{yy},\nonumber\\
&\tilde\beta_{zx}=-x W_{yx},\nonumber\\
&\tilde\beta_{zy}=\phi_{zy}-x W_{yy},
\end{align}
into the translational gauge field 
\begin{align}
\phi_{zy}=-\frac{\Omega}{2\pi}\,\kappa r K_1(\kappa r),
\end{align}
and the rotational gauge field
\begin{align}
\label{W-pot}
W_{yx}\equiv-\varphi^*_{yx},\qquad W_{yy}\equiv-\varphi^*_{yy}.
\end{align}
Thus, the negative disclination loop density~(\ref{bt-pl}) 
is equivalent to the rotational gauge potential~(\ref{W-pot}).

\section{Conclusion}
The field theory of elastoplasticity has been employed on the consideration
of a straight twist disclination in an infinitely extended body. 
We were able to calculate the elastic and plastic fields.
We found that the elastic stress, elastic strain, elastic bend-twist, 
dislocation density
and disclination density are 
continuous and the displacement, plastic distortion, rotation and 
the disclination loop density
of the twist disclination are discontinuous fields.
Exact analytical solutions for all characteristic field quantities of a 
twist disclination have been 
reported which demonstrate the elimination of ``classical'' singularities 
at the disclination line. 
The disclination core appears naturally as a result of 
the smoothing of the rotation vector profile.
In addition, we pointed out and discussed the relation between the twist
disclination with Frank vector $\frank\equiv (0,\Omega,0)$ and an edge
dislocation with Burgers vector $\burger\equiv(b,0,0)$.
We were able to calculate the effective Frank and Burgers vector of 
the twist disclination.
The force stress of a twist disclination calculated in the field theory of 
elastoplasticity agrees with the stress calculated within
the theory of nonlocal elasticity and strain gradient elasticity. 
The reason is that in all three theories the fundamental equation
for the force stress has the form of an inhomogeneous Helmholtz
equation (see Eq.~(\ref{stress-fe})).
The solutions of a twist disclination 
considered in this paper could be help in studies of
mechanical behaviour of nano-objects including
nanotubes and nanomembranes and of disclinated nanoparticles. 
Last but not least, using the geometric framework of $ISO(3)$-gauge theory
of defects we have found the translational and rotational gauge fields 
of a twist disclination. It turned out that the (negative) disclination
loop density is equivalent to the rotational gauge field. 
In general, the (negative) gauge fields may be considered as the plastic parts
in the field theory of elastoplasticity.

\subsection*{Acknowledgement}
The author is grateful to Drs.~Mikhail Yu.~Gutkin and Gerald Wagner
for some comments on this paper.
The author
acknowledges the Max-Planck-Institut f{\"u}r 
Mathematik in den Naturwissenschaften for financial support.

\end{document}